\documentclass[12pt]{iopart}

\usepackage{amsfonts}
\usepackage{amssymb}
\usepackage{mathrsfs}  
\usepackage{accents}
\usepackage{bm}
\usepackage{color}
\usepackage{graphicx}
\usepackage{amsthm}
\usepackage{enumerate}
\usepackage{mathrsfs}
\newlength{\dhatheight}
\newcommand{\mathcalligra}[1]{{\mathfrak #1}}

\usepackage{hyperref}
\usepackage{calligra}
\usepackage{relsize} 

\begin{document}

\title[Martingales and first-passage problems]{Martingale approach for first-passage problems of time-additive observables in Markov processes }

\author{Izaak Neri}

\address{Department of Mathematics, King’s College London, Strand, London, WC2R 2LS, UK
}
\ead{izaak.neri@kcl.ac.uk}

\begin{abstract}
We develop  a method based on martingales to study  first-passage problems of    time-additive observables exiting an interval of finite width in a Markov process.    In the limit that the interval width is large, we   derive generic expressions for the splitting probability and the cumulants of the first-passage time.   These expressions relate  first-passage  quantities to the large deviation properties of the time-additive observable.   We find that there are three qualitatively different regimes depending on the properties of the large deviation rate function of the time-additive observable. These regimes correspond to exponential, super-exponential, or sub-exponential suppression of events at the unlikely boundary of the interval.  Furthermore, we show that  the statistics of first-passage times at both interval boundaries are  in general different, even for symmetric thresholds and in the limit of large interval widths.   While   the statistics of the times to reach the  likely boundary are determined by the cumulants of the time-additive observables in the original process, those at the unlikely boundary are determined by 
a dual process.    We obtain these results from a one-parameter family of positive martingales that we call Perron martingales, as these are related to the Perron root of a tilted version of  the transition rate matrix defining  the  Markov process.   Furthermore, we show that each eigenpair of the tilted matrix has a one-parameter family of martingales.   To solve first-passage problems at finite thresholds, we generally require all one-parameter families of martingales, including the non-positive ones. We illustrate this by solving the first-passage problem for run-and-tumble particles exiting an interval of finite width.
\end{abstract}

\section{Introduction}
The gambler's ruin problem is a first-passage problem of a random walker with two absorbing boundaries.   In the 17th century,  Pascal introduced this question  shortly after his communication  with Fermat on the problem of points~\cite{devlin2010unfinished}, and it appeared in print  for the first time  in Christiaan Huygens' treatise entitled ``{\it Van Rekeningh in Spelen van Geluck}" (On calculations for games of chance) \cite{huygens}.

\begin{figure}[h!]\centering
{\includegraphics[width=0.48\textwidth]{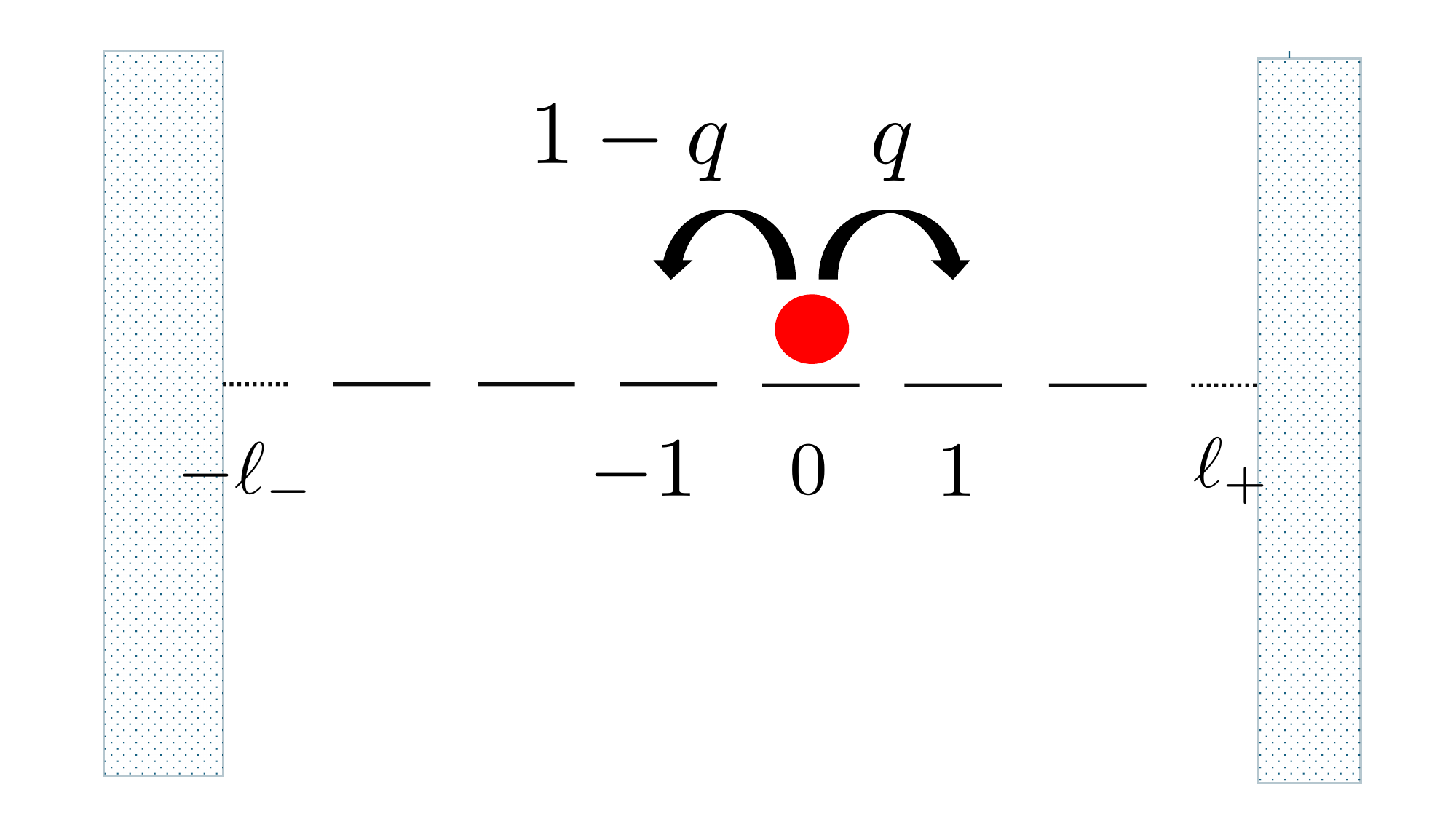}}
\includegraphics[width=0.5\textwidth]{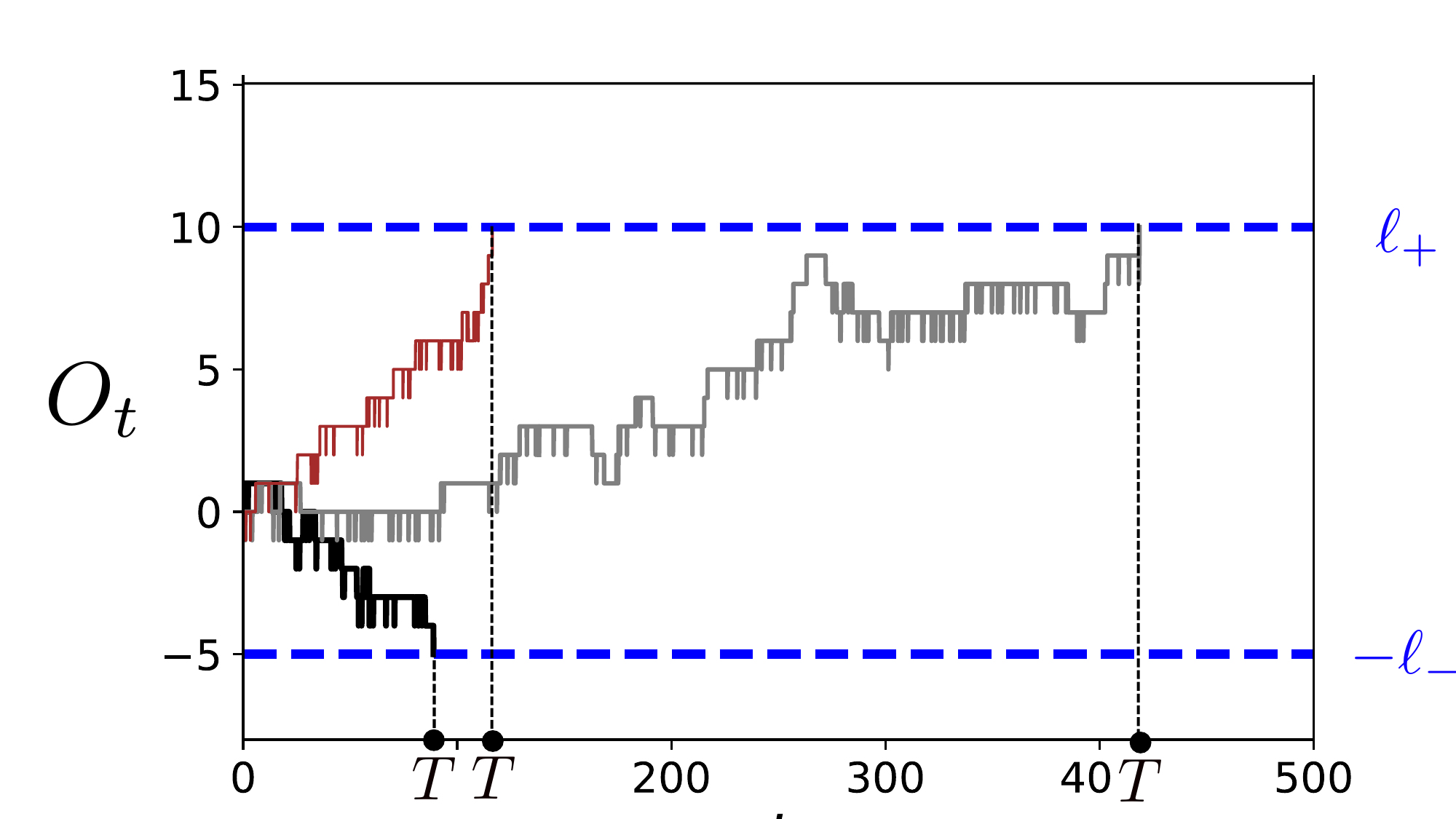}
\caption{Left: Illustration of the  standard version of Gambler's ruin problem.  Right: Gambler's ruin problem for a time-additive observable $O_t$  in a Markov jump process.   }\label{fig1}   
\end{figure}  

In  text books gambler's ruin problem is often formulated as follows (e.g., see  Chapter XIV of Ref.~\cite{feller1966introduction}).      Consider a particle that moves on a one-dimensional lattice of  finite length.    Its position denotes   the stake of a gambler that plays  a game of chance against another player.     The particle moves in discrete time steps with a probability $q$ to the right and with a probability $1-q$ to the left  (see  left panel of Fig.~1 for an illustration).   The process terminates as soon as the particle reaches one of the two end points of the lattice, corresponding with one of the two gamblers losing the entirety of their initial stake.    The main quantities of interest are the probability  that the process terminates at the  left (or equivalently the right) threshold,  and  the statistics of  the total duration of the  process.

Here, we consider a generalisation of  gambler's ruin problem that applies to  time-additive observables, $O_t$, in Markov processes, $X_t$, where $t\geq 0$ is a continuous time index.     A time-additive observable is a real-valued observable whose difference $O_t-O_{s}$  is fully determined by the trajectory of $X$ in the time interval $[s,t]$~\cite{touchette2009large, harris1, touchette2018introduction,  lazarescu2015physicist, monthus2021large, monthus2022microcanonical}.  Examples of time-additive observables are, amongst others,  energy and particle fluxes, the distance traversed by  a random walker, or the time   a process spends in a certain state.  
The first passage problem for  $O_t$, similar to the gambler's ruin problem, involves calculating the statistics for the times  $T$ when $O_t$  first exits the open interval $(-\ell_-,\ell_+)$, defined as
\begin{equation}
T := {\rm min}\left\{t\geq 0: O_t \notin (-\ell_-,\ell_+)\right\}, \label{eq:gamblerRuin}
\end{equation}
and calculating the splitting probabilities $p_-$ ($p_+$) that $O_t$ exits the interval from the negative (positive) side.  
We assume, without loss of generality, that  $\overline{o} := \lim_{t\rightarrow \infty}\langle O_t\rangle/t \geq 0$, where $\langle \cdot\rangle$ denotes an average over many realisations of the process $X_t$ and that $O_0=0$; if $\overline{o}<0$, then we can study the equivalent first-passage problem for $-O_t$.     Hence, the events at the negative threshold are suppressed, except when $\overline{o}=0$.

The first-passage problem (\ref{eq:gamblerRuin})   can be seen as a generalisation of    first-passage problems studied in statistical physics, such as escape problems of  particles (photons or neutrons) that move through a disordered medium~\cite{RevModPhys.89.015005, klinger2022splitting}, motor proteins   that deliver cargoes to the two  end points of a biofilament~\cite{muller2008tug, newby2010local, newby2010random},   or  self-propelled particles that escape a bounded region of space~\cite{malakar2018steady, gueneau2024relating, gueneau2024run}.     For mathematical modelling, first-passage problems of time-additive observables are useful  in various areas in science whenever we can observe the timing of discrete events.  Examples are forward/backward stepping of molecular motors along a biofilament~\cite{fox2001rectified, kolomeisky2005understanding},  the decision times of humans or animals in perceptual decision tasks~\cite{ratcliff2008diffusion, kira2015neural}, or  the timing of decisions by cells~\cite{siggia2013decisions,desponds2020mechanism}.

First-passage problems of time-additive observables have also been studied in nonequilibrium thermodynamics as a probe for   nonequilibrium fluctuations~\cite{decision1, saito2016waiting, neri2017statistics, gingrich2017fundamenta,  falasco2020dissipation, pal2021thermodynamic, hiura2021kinetic, wampler2021skewness,  neri2022universal, neri2022estimating}.    In this case, the time-additive observables are often   fluctuating currents, $J_t$, such as, the total   charge that has been transported through an electrical wire, or the net number of times that a chemical reaction has been completed; although for thermodynamic  purposes it is also interesting to study time-additive observables that are not   currents, such as,  the work done on a system of interest~\cite{mamede2024work} or the dynamical activity~\cite{garrahan2017simple}.     In the limit of large thresholds the fluctuations of the first-passage times of currents are constrained  by the rate of dissipation, as expressed by trade-off inequalities of the form 
\begin{equation}
\dot{s} \geq \frac{\alpha }{\epsilon_{\rm unc}\langle T\rangle} (1+\mathcalligra{o}_{\ell_{\rm min}}(1)) \label{eq:thermo1}
\end{equation} 
where $\dot{s}$ is the rate of dissipation, i.e., the amount of entropy produced per unit of time in natural units of information (nats); where $\alpha$ is a constant prefactor;  where $\langle T\rangle$ is the average of  the first-passage time $T$ over many realisations of the process;  where $\epsilon_{\rm unc}$ is a dimensionless  measure of uncertainty; where $\mathcalligra{o}_{\ell_{\rm min}}(1)$ denotes an arbitrary function that converges to zero when $\ell_{\rm min} = {\rm min}\left\{\ell_-,\ell_+\right\}$ diverges.   Examples are the thermodynamic uncertainty relation for first-passage times~\cite{gingrich2017fundamenta}, in which case 
\begin{equation}
\alpha = 2 \quad {\rm and} \quad \epsilon_{\rm unc} = \frac{\langle T^2\rangle-\langle T\rangle^2}{\langle T\rangle^2}, \label{eq:thermo2}
\end{equation}
and a splitting probability version of the thermodynamic uncertainty relation with now~\cite{decision1,neri2022universal, raghu2024effective} 
\begin{equation}
\alpha = \frac{\ell_+}{\ell_-} \quad {\rm and} \quad  \epsilon_{\rm unc} = \frac{1}{ |\ln p_-|}. 
\end{equation}
The interest of these trade off relations is in their appealing physical  interpretation, and the fact that they hold generically for fluctuating currents  in Markov jump processes.

In this Paper, we develop a method based on martingales to calculate the properties of the first-passage time $T$ of a generic time-additive observable $O_t$ in a Markov process $X_t$.     The martingale  approach for first-passage problems  dates back  to the  work of Jean Ville~\cite{Ville} and is an alternative to the "classical" approach  based on solving a Fokker-Planck equation with absorbing boundary conditions~\cite{feller1966introduction,redner2001guide}.   Following the pioneering work of Jean Ville, martingales have been used to study first-passage problems for various stochastic processes, such as,    biased random walkers~\cite{neri2022universal, roldan2022martingales},   one-dimensional time-homogeneous driven diffusions~\cite{sarmiento2024area}, and  driven diffusions that have time-dependent parameters~\cite{lin1998double, srivastava2017martingale}.   However, to the best of our knowledge, the martingale method has not been presented  before  for  generic time-additive observables in Markov processes.  A presentation of this general theory, as we do here, reveals interesting connections between  martingales, first-passage problems, and large deviation theory.  

In the generic approach that we develop here, 
we associate to each time-additive observable $O_t$ multiple one-parameter families of martingales.  Each one-parameter family is   associated with a distinct eigenvalue of the, so-called, tilted transition-rate matrix that appears in the theory of  large deviations~\cite{touchette2009large, chetrite_nonequilibrium_2015}.   The Perron root of the tilted matrix  determines the scaled cumulant generating function of the time extensive observable $O_t$ in the limit of $t\gg 1$, and we call the corresponding martingale the    {\it Perron martingale}.  As we show in this Paper, the Perron martingale relates the statistics of the first-passage times $T$ in the limit of large thresholds to  the  large deviation  properties of $O_t$.  
At finite thresholds, Perron martingales alone typically do not suffice to fully characterize the statistics of $T$. Nevertheless, first-passage problems can often be solved when using both the Perron martingale and the martingales associated with the other eigenvalues of the tilted generator,  as we demonstrate in this Paper for the first-passage problem of a run-and-tumble particle that escapes an interval of finite width.

The paper is structured as follows.  In  Sec.~\ref{sec:setup} we define the problem of interest, viz.,  first-passage problems of  time-additive observables with two absorbing boundaries.  For clarity, the  following sections, Secs.~\ref{sec:martingale} to \ref{sec:LDPMean}, start with a subsection presenting their main results, followed by subsections with the derivations of the results.
 In Sec.~\ref{sec:martingale}, we introduce   one-parameter families of martingales associated with time-additive observables, including the Perron martingale.   These martingales constitute the main tool that we use throughout this paper to study  first-passage problems of time-additive observables.  
 In Sec.~\ref{sec:moment}, we use  Perron martingales to determine the statistics of   first-passage times   in the limit of large thresholds and for nonzero average rates $(
\overline{o}\neq 0 )$.   In Sec.~\ref{sec:thermoBound}, we use the results from Sec.~\ref{sec:moment} to derive a  generic thermodynamic bound on the statistics of  first-passage times of fluctuating currents at the negative threshold, extending previous results that apply at the positive threshold~\cite{gingrich2017fundamenta}.  In  Sec.~\ref{sec:moment2}, we  relate the cumulants of $T$ in the limit of large thresholds to the cumulants of $O_t$.   Interestingly, we find that the cumulants at the negative threshold are different from those at the positive threshold.      In Sec.~\ref{sec:finite}, we   derive  for  a certain class of time-additive observables the  splitting probabilities and the  generating functions of $T$ at finite values for the thresholds.    In Sec.~\ref{sec:LDPMean} we study  the statistics of $T$ for  time-additive observables that have zero average values ($\overline{o}=0$), and we show that in this case the first-passage problem exhibits diffusive behaviour.    In the next two sections, Secs.~\ref{sec:randomwalker} and \ref{sec:runAndTumble},  we use martingales to solve  first-passage problems of active particles that escape from a bounded region of space.  In Sec.~\ref{sec:randomwalker} we consider a biased random walker escaping an infinitely long strip of finite width, and we  show  that in this example the  cumulants of $T$ at the two thresholds are distinct.   In Sec.~\ref{sec:runAndTumble} we consider a one-dimensional run-and-tumble particle escaping from an interval  on the real line.     In this  example  the Perron martingales do not suffice to determine the splitting probabilities and mean first-passage time.   Nevertheless, we solve this problem  by considering a second family of nonpositive martingales associated with the particle's position, which illustrates the  martingale method on a more complicated example.  We end the paper with a discussion in Sec.~\ref{sec:discussion} and a couple of appendices with technical details.

\section{System setup: first-passage problems of time-additive observables}  \label{sec:setup}
\subsection{Markov jump processes}
We consider   time-homogeneous Markov jump processes $X_t$ that take values in a finite set  $\mathcal{X}$  and with $t\geq 0$ a continuous time index.   Time-homogeneous Markov jump processes are fully specified by their transition rate matrix $\mathbf{q}$  and the probability mass function $p_{X_0}(x)$ of $X_t$ at the initial time $t=0$~\cite{norris1998markov, liggett2010continuous}.  The  off-diagonal entries $\mathbf{q}_{xy}$ of  $\mathbf{q}$  denote the rates at which $X_t$ jumps from $x$ to $y$, and the         diagonal entries $\mathbf{q}_{xx} = -\sum_{y\in \mathcal{X}\setminus \left\{x\right\}}\mathbf{q}_{xy}$ denote the  rates at which the process leaves  the state $x$ (times minus one).         

The probability mass function $p_t(x)$ of $X_t$ solves the differential equation 
\begin{equation}
\partial_t p_t(x) =  \mathcal{L}^\ast[p_t(x)] = \sum_{y\in \mathcal{X}}\mathbf{q}_{yx}p_t(y)
\end{equation}
with  initial condition $p_{X_0}(x) = p_0(x)$.    
We assume that $X_t$ is ergodic so that there exists a unique probability mass function $p_{\rm ss}(x)$ for which $\mathcal{L}^\ast[p_{\rm ss}] = 0$~\cite{bremaud2013markov}.    For Markov jump processes defined on a finite set  $\mathcal{X}$, a sufficient condition for ergodicity is that the graph $\mathcal{G} = (\mathcal{X},\mathcal{E})$ of admissible transitions is strongly connected, where 
\begin{equation}
\mathcal{E} =  \left\{(x,y)\in \mathcal{X}^2: \mathbf{q}_{xy}>0   \right\}
\end{equation}
is the set of directed links of $\mathcal{G}$.  

We denote by $\langle \cdot \rangle$ averages over multiple realisations of the process $X_t$  with the statistics of $X$ described  by $\mathbf{q}$; the corresponding probability measure is denoted by $\mathbb{P}$.

\subsection{Time-additive observables}
We consider {\it time-additive observables} of the form
\begin{equation}
O_t := \sum_{x\in\mathcal{X}}c_x N^x_t+ \sum_{(x,y)\in \mathcal{E}}c_{x y}N^{x y}_t   \label{eq:generic}
\end{equation}
where $N^x_t$ is the amount of  time that $X$  has spent in the state $x$ in the interval $[0,t]$, and where
$N^{xy}_t$ counts  the number of jumps from $X_{s^-} = x$ to $X_{s} = y$ with $s\in[0,t]$ and $s^- = \lim_{\epsilon\uparrow 0}(s-\epsilon)$.      

Since $X_t$ is an ergodic process the average rate of change in $O_t$, denoted by $\overline{o}$, takes the expression
\begin{equation}
\overline{o} := \lim_{t\rightarrow \infty} \frac{\langle O_t\rangle}{t} =  \sum_{x\in \mathcal{X}}c_x \: p_{\rm ss}(x)+ \sum_{(x,y)\in \mathcal{E}}  c_{xy} \: p_{\rm ss}(x) \mathbf{q}_{xy}  . \label{eq:overlineo}
\end{equation}
 We assume, without loss of generality, that $\overline{o}\geq 0$ (if $\overline{o} <0$, then we can consider $-O_t$ instead of $O_t$).

{\it Fluctuating currents}, $J_t$, are time-additive observables for which $c_x=0$ and  $c_{xy}=-c_{yx}$ for all $x,y\in\mathcal{X}$.    Energy  and particle fluxes are examples of fluctuating currents.  In nonequilibrium thermodynamics,   fluctuating currents are of particular interest, as   average currents,  
\begin{equation}
\overline{j} := \lim_{t\rightarrow \infty} \frac{\langle J_t\rangle}{t}, 
\end{equation} 
are, in general, nonzero  far from  thermal equilibrium, while $\overline{j}=0$ for systems in thermal equilibrium.    Therefore, currents with nonzero average values are a hallmark of nonequilibrium physics.  

The  time-additive observable $O_t$ obeys a large deviation principle~\cite{maes2008canonical, maes2008steady,  touchette2009large, dembo2009large, jack2010large, Bert2015, bertini2015flows,  carugno2022graph}, which implies that the rate function 
\begin{equation}
\lim_{t\rightarrow \infty} \frac{\ln p_{O_t/t}(o)}{t} =: -\mathcal{I}(o)
 \end{equation}
 exists.   Here, we used the notation $p_{O_t/t}$ for the probability distribution of $O_t/t$.    According to the G\"{a}rtner-Ellis theorem (see Theorem 2.3.6 in \cite{dembo2009large}), the rate function $\mathcal{I}(o)$ is the Fenchel-Legendre transform of the {\it scaled cumulant  generating function}   
 \begin{equation}
 \lambda_O(a) := \lim_{t\rightarrow \infty} \frac{1}{t}\ln \langle \exp(-a O_t) \rangle,   \label{def:lambdaO}
 \end{equation}
 such that 
 \begin{equation}
\mathcal{I}(o) =  {\rm max}_{a\in \mathbb{R}}(-\lambda_O(a)-ao). 
 \end{equation}
Expanding $\lambda_O(a)$ around $a=0$, we get 
\begin{equation}
\lambda_O(a) = -\overline{o} \: a + \sigma^2_O  \: \frac{a^2}{2} + \mathcal{O}(a^3) ,\label{eq:taylor}
\end{equation}
where $\overline{o}$ is the average rate defined in Eq.~(\ref{eq:overlineo}),  where $\sigma^2_O$ is the diffusivity coefficient of  the observable $O$, 
\begin{equation}
\sigma^2_O := \lim_{t\rightarrow \infty}  \frac{\langle O^2_t\rangle - \langle O_t\rangle^2}{t},\label{eq:diffusivity}
\end{equation}
and where $\mathcal{O}(\cdot)$ is the big-O notation.  
The higher order coefficients  in the Taylor series  (\ref{eq:taylor}) generate the higher order central moments of  $O$ rescaled by~$t$.    

The scaled cumulant generating function $\lambda_O(a)$ is   the Perron root of the {\it tilted matrix} $\tilde{\mathbf{q}}(a)$  whose elements are given by~\cite{touchette2009large,  lebowitz1999gallavotti, baiesi2009computation, jack2010large, chetrite_nonequilibrium_2015}
\begin{equation}
\tilde{\mathbf{q}}_{xy}  := \left\{\begin{array}{ccc} e^{-ac_{xy}}\mathbf{q}_{xy}, &{\rm if}& x\neq y,\\ \mathbf{q}_{xx} -a c_x, &{\rm if}&  x=y.\end{array}\right. \label{eq:qtilde}
\end{equation}
Hence, we can obtain $\lambda_O(a)$ from diagonalising the tilted matrix  $\tilde{\mathbf{q}}(a)$.

\subsection{First-passage problems with two thresholds}
In this Paper, we examine first-passage problems for time extensive observables similar to the gambler's ruin problem. Specifically, we make a study of processes with a finite termination time, defined by the {\it first-passage time} 
\begin{equation}
T := {\rm min}\left\{t\geq 0 : O_t\notin (-\ell_-,\ell_+)\right\} \label{eq:T}
\end{equation} 
when $O$ exits the finite open interval $(-\ell_-, \ell_+)$.

The  {\it splitting probabilities} $p_-$ and $p_+$ are the probabilities that the process terminates at the negative and positive thresholds, respectively, i.e., 
\begin{equation}
p_- := \mathbb{P}\left(O_T\leq -\ell_-\right) \quad {\rm and} \quad p_+ := \mathbb{P}\left(O_T\geq \ell_+\right). 
\end{equation}
Note that $p_-$ is the analogue of the  probability  of the gambler's ruin~\cite{feller1966introduction}.
For processes with $\overline{o}\neq 0$, it holds with probability one that  $T<\infty$, and therefore 
\begin{equation}
p_-+p_+=1.
\end{equation}
If $\ell_+\rightarrow\infty$, then Eq.~(\ref{eq:T}) defines a first-passage problem with one absorbing boundary, and in such instances  $p_+$ is called the persistence or survival probability~\cite{bray2013persistence}.   
 
The statistics of $T$ are determined by  their {\it moment generating functions} 
\begin{equation}
g_+(\mu) := \langle e^{\mu T} \rangle_+ \quad {\rm and} \quad  g_-(\mu) := \langle e^{\mu T} \rangle_-, \label{eq:genFunc}
\end{equation}
where $\langle \cdot \rangle_+ = \langle \cdot | O_T \geq \ell_+ \rangle $ and  $\langle \cdot \rangle_- = \langle \cdot | O_T \leq -\ell_- \rangle $ are expectation values conditioned on events terminating  at the positive and negative threshold, respectively.  

\subsection{Rate of dissipation and fluctuating entropy production} 
We define two quantities of interest for the physics of nonequilibrium systems.   The {\it rate of dissipation},  $\dot{s}$, quantifies the number of nats  of entropy produced in the environment per unit of time~\cite{bookPrigo}.  For Markov jump processes,  the rate of dissipation admits the expression~\cite{schnakenberg}  
\begin{equation}
\dot{s} := \sum_{(x,y)\in \mathcal{E}} p_{\rm ss}(x)\mathbf{q}_{xy} \ln \frac{p_{\rm ss}(x)\mathbf{q}_{xy}}{p_{\rm ss}(y)\mathbf{q}_{yx}}. \label{eq:sdotDef}
\end{equation}
Notice that since we use the natural logarithm in the definition of $\dot{s}$, entropy is  measured in natural units of information (nats). 

  The {\it fluctuating entropy production}, $S_t$, is defined through a change of measure.    Specifically,  let 
 $\mathbb{P}^\dagger$ be the probability  measure associated with the time-reversed process with transition rates 
\begin{equation}
\mathbf{q}^\dagger_{xy} := \left\{\begin{array}{ccc} \mathbf{q}_{yx} \frac{p_{\rm ss}(y)}{p_{\rm ss}(x)},&{\rm if}& x\neq y, \\  \mathbf{q}_{xx}, &{\rm if}& x=y, \end{array} \right.
\end{equation}
then 
$S_t$ is the process for which  
\begin{equation}
\langle   e^{-S_t} f(X_{[0,t]}) \rangle_{\mathbb{P}} = \langle  f(X_{[0,t]})\rangle_{\mathbb{P}^\dagger},
\end{equation}
holds for all bounded functions $f$ defined on the trajectories $X_{[0,t]} := \left\{X(s):s\in [0,t]\right\}$ of the process $X$.        For Markov jump processes, $S_t$ takes the  form 
\begin{equation}
S_t  = \sum_{(x,y)\in \mathcal{E}} N^{xy}_t  \: \ln \frac{p_{\rm ss}(x)\mathbf{q}_{xy}}{p_{\rm ss}(y)\mathbf{q}_{yx}} ,  \label{eq:St}
\end{equation} 
and since $X_t$ is ergodic, it holds that 
\begin{equation}
\dot{s} = \lim_{t\rightarrow \infty}\langle S_t\rangle/t . 
\end{equation}

\section{Martingales associated with time-additive observables}  \label{sec:martingale}
For each time-additive observable, $O_t$, we define  multiple  one-parameter families of   martingales  $M_t$.     A martingale $M_t$ relative to $X_t$ is a process that is driftless~\cite{williams1991probability}, i.e.,
\begin{equation}
\langle M_t|X_{[0,s]}\rangle = M_s \label{eq:driftless}
\end{equation}
for all values $s\in[0,t]$.   A sufficient condition for the existence of the conditional expectation in  the left-hand side of Eq.~(\ref{eq:driftless}) is that  $\langle |M_t|\rangle<\infty$ (see Theorem 10.1.1 in~\cite{Dudley}).

\subsection{Main results}
Let $\mu_O(a)$ be an  eigenvalue of the tilted matrix $\tilde{\mathbf{q}}(a)$ defined in Eq.~(\ref{eq:qtilde}), and let $\zeta_a(x)$  be the right eigenvector of  $\tilde{\mathbf{q}}$ associated with $\mu_O(a)$, i.e.,
\begin{equation}
\sum_{y\in \mathcal{X}}\tilde{\mathbf{q}}_{xy}\zeta_a(y) = \mu_O(a)\zeta_a(x), \quad \forall x\in \mathcal{X}. \label{eq:spectralProblem}
\end{equation}
It then holds that  the processes
\begin{equation}
M_t := \zeta_a(X_t)\exp\left(-aO_t -\mu_O(a) t\right), \label{eq:Mart2}
\end{equation}
 are martingales for all values of $a\in \mathbb{R}$.
 
 Notice that if $\tilde{\mathbf{q}}(a)$ is a normal matrix, then the spectral problem (\ref{eq:spectralProblem}) has $|\mathcal{X}|$ linearly independent solutions, and thus we got  $|\mathcal{X}|$ martingales.   However, in general we expect that the number of martingales is smaller than that, as nonsymmetric matrices are not guaranteed to be normal.   The  martingales are real-valued when  the eigenvalue $\mu_O$ is real.

A specifically important class of martingales of the form (\ref{eq:Mart2}) are those for which   $\mu_O(a)$ equals the Perron root  of $\tilde{\mathbf{q}}$, which is also the scaled cumulant generating function  $\lambda_O(a)$ defined in (\ref{def:lambdaO}).   We call these martingales {\it Perron martingales}, and they take the form
\begin{equation}
M_t = \phi_a(X_t)\exp\left(-aO_t - \lambda_O(a)t\right), \label{eq:Mart}
\end{equation}
where $\phi_a(x)$ is the right eigenvector of the Perron root $\lambda_O(a) $ of $\tilde{\mathbf{q}}$.     A distinction between the  martingales (\ref{eq:Mart2}) and (\ref{eq:Mart}) is that the latter are positive (because of the Perron-Frobenius theorem all entries of $\phi_a(x)$ have the same sign~\cite{dembo2009large}), while the former are not necessarily positive.

In this Paper, we mainly focus on the Perron martingale (\ref{eq:Mart}) as it relates the large deviations of $O_t$ (for large $t$) with the large deviations of the   first-passage time $T$ (for large thresholds $\ell_+$ and $\ell_-$).    Nevertheless,  the  nonpositive martingales (\ref{eq:Mart2}) are important at finite thresholds, and  we demonstrate this in  Sec.~\ref{sec:runAndTumble} by solving the first-passage problem of a run-and-tumble particle.

\subsection{Derivation}
We show that the $M_t$ defined in (\ref{eq:Mart2}) are martingales.

Since the pair $(X_t,O_t)$ is a Markov process, it is sufficient that the process is locally driftless, i.e., 
\begin{equation}
\lim_{t\rightarrow 0} \frac{\langle f_t(X_t,O_t)| X_0=x,O_0=o \rangle -f_t(x,o)}{t} = 0 \label{eq:Deriv}
\end{equation} 
where 
\begin{equation}
f_t(x,o) = \zeta_a(x)\exp(-ao - \mu_O(a)t) .\label{eq:f}
\end{equation}
Since $(X_t,O_t)$ is a Markov process, the  left-hand side of Eq.~(\ref{eq:Deriv}) can be expressed as 
\begin{equation}
\partial_t f_t(x,o)  + \mathcal{L}[f_t](x,o)   = 0 \label{eq:f2}
\end{equation}
where $\mathcal{L}$ is the generator of the joint process  $(X_t,O_t)$ that acts on  (bounded) functions $g(x,o)$  as follows 
\begin{equation}
\mathcal{L}[g](x,o) = \sum_{y\in \mathcal{X}\setminus \left\{x\right\}} \mathbf{q}_{xy}  \left[g(y,o+c_{x y})  - g(x,o) \right]  +   c_x \partial_o g(x,o) .  \label{eq:Gen}
\end{equation}  
  Applying the generator $\mathcal{L}$ to the function $f$ of (\ref{eq:f}), and using that    $\partial_t f_t = -\mu_O(a) f_t$, $\partial_of_t = -a f_t$, and $f_t(x,o+c_{xy}) = f_t(x,o)\exp(-ac_{xy})$, we find that Eq.~(\ref{eq:f2})  holds when
\begin{equation}
 \sum_{y\in \mathcal{X}}  \tilde{\mathbf{q}}_{xy}(a) \zeta_a(y) = \mu_O(a) \zeta_a(x).  \label{eq:phiHold}
\end{equation}  
Equation (\ref{eq:phiHold}) holds for all eigenpairs $(\mu_O,\zeta_a)$ of $\tilde{\mathbf{q}}(a)$, and therefore indeed the processes  (\ref{eq:Mart2}) are martingales.

If $M_t$ is a Perron martingale, then $M_t$ is a positive random variable, and thus a Radon-Nikodym density process (see Sec.9.3 of Ref.~\cite{avanzini2024methods}).   Indeed, as shown in Eq.~(119) of Ref.~\cite{chetrite_nonequilibrium_2015}, Eq.~(\ref{eq:Mart}) can be expressed as a Radon-Nikodym density process.

\subsection{Examples}
We consider  a few examples of   Perron martingales of the form (\ref{eq:Mart}), including  cases studied before in the literature.

If the time-additive observable is a fluctuating current, $O=J$, and if in addition  $a=a^\ast$, with $a^\ast$ the nonzero root of 
\begin{equation}
\lambda_J(a^\ast) = 0,
\end{equation} 
then the Perron martingale (\ref{eq:Mart}) reads 
\begin{equation}
M_t = \phi_{a^\ast}(X_t) \exp(-a^\ast J_t), \label{eq:MartSP}
\end{equation}
which is the same martingale as studied in Ref.~\cite{raghu2024effective}. The quantity $a^\ast$ is called the {\it effective affinity}, as it extends  properties of thermodynamic affinities of uncoupled currents to systems with coupled currents, see Ref.~\cite{raghu2024effective}.     

In the special case that the fluctuating current is the stochastic entropy production $S_t$, as defined in (\ref{eq:St}), it holds that  $a^\ast=1$ and $\phi_{1}(x)=1$ (see~\ref{app:entropy}), and thus
the martingale (\ref{eq:MartSP})  is the exponentiated negative entropy production~\cite{chetrite2011two, neri2017statistics, neri2019integral} 
\begin{equation}
M_t= \exp(-S_t). \label{eq:MartSPS}
\end{equation} 
 Another special limiting case is when $J=J^{xy}:= N^{xy}_t-N^{yx}_t$ equals an edge current, in which case (\ref{eq:MartSP}) is equivalent with the martingale in~\cite{neri2023extreme}.

In the case of a unicyclic Markov process, we determine the Perron martingale  (\ref{eq:Mart}) for an arbitrary fluctuating current.  
Consider  the process  $X_t\in \mathcal{X}=\left\{1,2,\ldots,n\right\}$ with the transition rate matrix 
\begin{equation}
\mathbf{q}_{xy} = k_+ \delta_{y,x+1} + k_- \delta_{y,x-1} - (k_-+k_+)\delta_{x,y} , \quad x,y\in \mathcal{X},
\end{equation}
where the $\delta$'s are Kronecker delta functions.    This process represents a particle hopping on a  ring of length $n$.  We implement periodic boundary conditions by using addition and subtraction in modulo $n$ ($n+1=1$ and $0=n$).  Let us first consider the Perron martingale of  the fluctuating current 
\begin{equation}
J^{\rm uni}_t =   \sum^{n}_{x=1} (N^{x (x+1)}_t-N^{(x+1)x}_t)
\end{equation}
 that measures the distance traversed by  $X_t$ along the ring.  The first passage problem of $J^{\rm uni}_t$ is also the first-passage problem of  a  biased random walker  on $\mathbb{Z}$.       In this case,   the Perron martingale reads (see Appendix D of Ref.~\cite{neri2022universal})
\begin{equation}
M_t = \exp\left(-aJ^{\rm uni}_t -[(e^{-a}-1)k_+ +(e^{a}-1)k_-]  t\right), \label{eq:MTRWLK}  
\end{equation}
where in the exponent of the right-hand side of (\ref{eq:MTRWLK})
 we recognise  the scaled cumulant generating function 
\begin{equation}
\lambda_{J^{\rm uni}}(a) = (e^{-a}-1)k_+ + (e^{a}-1)k_-. \label{eq:RWLambda}  
\end{equation}  
We can also express the martingale (\ref{eq:MTRWLK}) as
\begin{equation}
M_t  = \phi_a(X_t) \exp\left(-aJ_t -[(e^{-a}-1)k_+ +(e^{a}-1)k_-]  t\right), \label{eq:mexpr}
\end{equation}
where 
\begin{equation}
J_t =  \sum^{n}_{x=1} c_{x(x+1)}(N^{x (x+1)}_t-N^{(x+1)x}_t)
\end{equation}
is an arbitrary current with the cycle coefficient $\sum^{n}_{x=1}c_{x(x+1)}=n$,   and 
where 
\begin{equation}
\phi_a(x) = \alpha \:  \exp\left(a\sum^{x-1}_{y=1}c_{y(y+1)}- a  \: x\right).
\end{equation}
Here  $\alpha$  is a proportionality constant that sets the  sum $\sum^n_{x=1}\phi_a(x) = n$.    It follows from (\ref{eq:mexpr}) that   $M_t$ is the Perron martingale of an arbitrary  current $J_t$ for which it holds that   $ \sum^{n}_{x=1}c_{x(x+1)}=n$.    If the  coefficients $c_{x(x+1)}$ are non-identical, then  the   prefactor $\phi_a(x)$ is nonconstant.

\begin{figure}[h!]\centering
{ \includegraphics[width=0.3\textwidth]{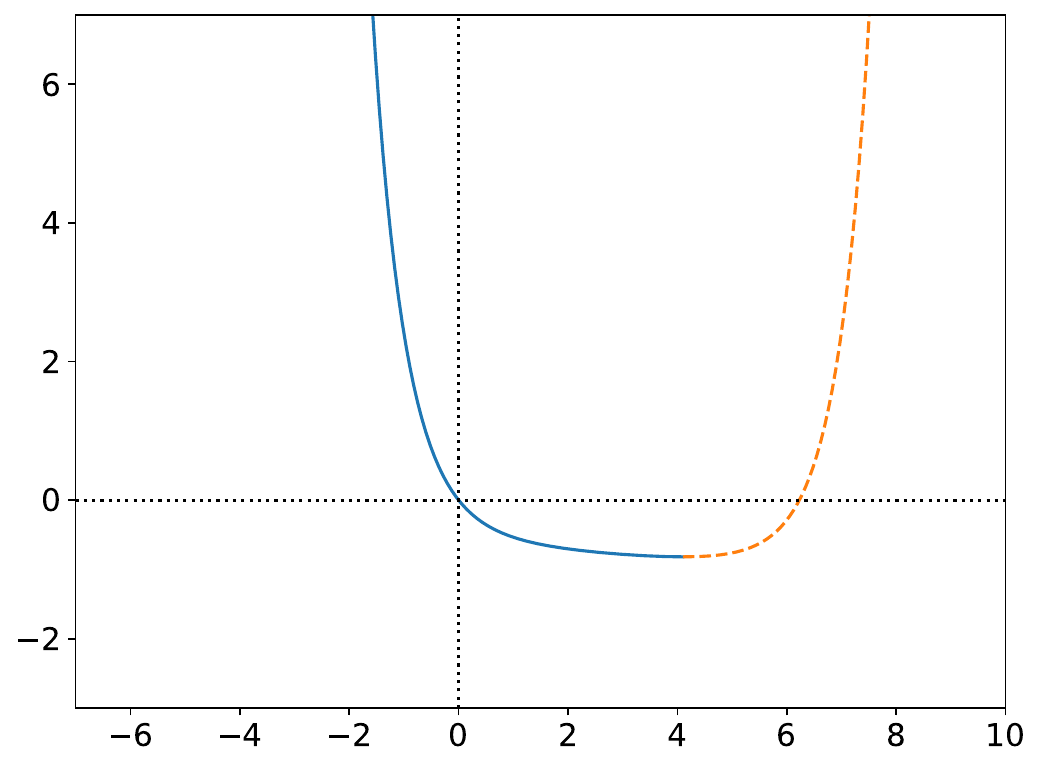}\put(-150,50){\small $\lambda_O$} 
\put(-68,-8){\small  $a$}}\hspace{0.4cm}
 { \includegraphics[width=0.31\textwidth]{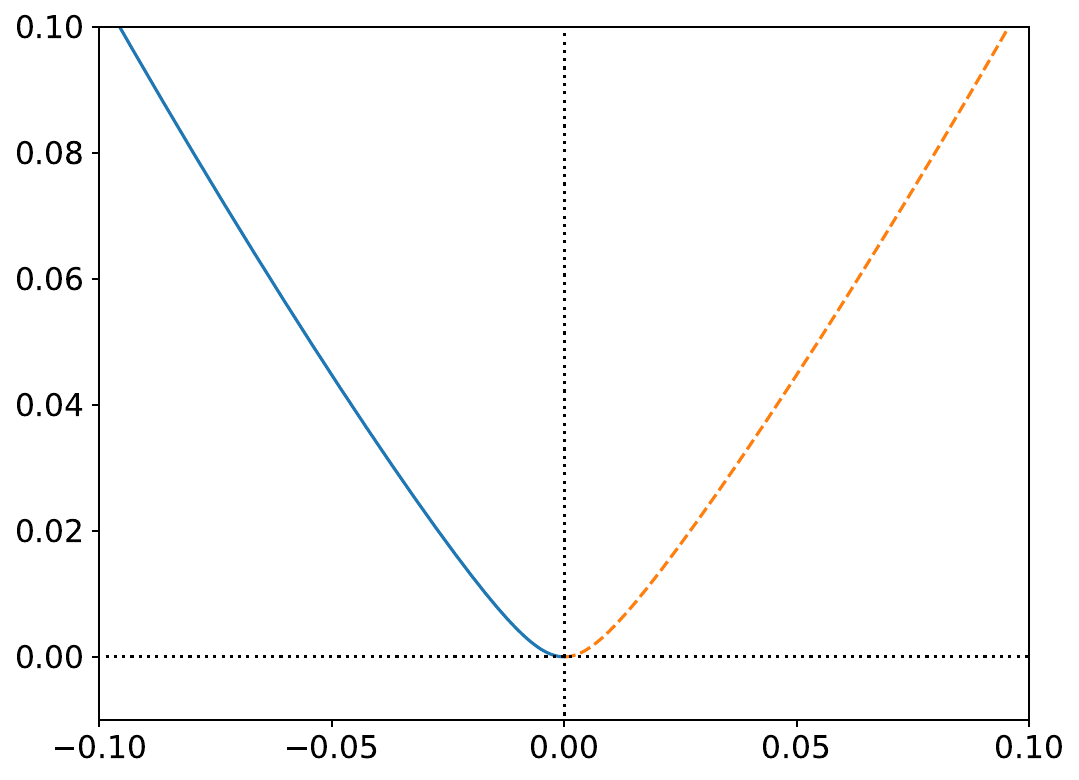}\put(-150,50){\small  $\lambda_O$} 
\put(-68,-8){\small  $a$}}
\hspace{0.4cm}
 {  \includegraphics[width=0.3\textwidth]{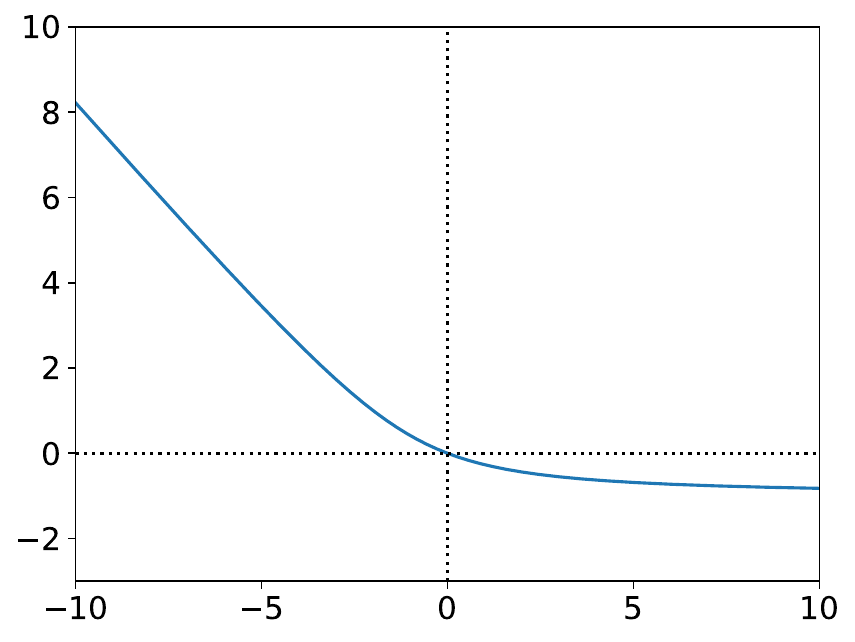}\put(-150,50){\small  $\lambda_O$} \put(-68,-8){\small  $a$}}
\caption{Scaled cumulant generating functions $\lambda_O(a)$ for  three qualitatively different cases corresponding with a time-additive observable $O$ that  is (i)  a fluctuating current  $J$ with $\overline{j}> 0$ (Left Panel); (ii) a fluctuation current $J$   with $\overline{j}=0$ (Middle Panel); and (iii) the time that the process $X$ spends in a certain state (Right Panel).     In these cases, the scaled cumulant generating function has the following  qualitative properties: (i) Equation~(\ref{eq:roots}) admits two solutions, and the minimum value of $\lambda_O(a)$ is nonzero (Left Panel);  (ii) Equation~(\ref{eq:roots}) admits two solutions, and the minimum value of $\lambda_O(a)$ equals zero (Middle Panel); and (iii)   Equation~(\ref{eq:roots}) admits one solution (Right Panel).    
   The specific $\lambda_O(a)$ functions plotted are the following: (i) plot of $\lambda_J$ given by Eq.~(\ref{eq:lambdaJ2D})   for  $\nu=5$, $\rho=2$, and $\Delta = 0.7$ (Left Panel);    (ii) plot of  $\mu_+$ given by Eq.~(\ref{eq:muPMRT}) for parameters $k_{\rm b}=1$, $k_{\rm f}=2$, and $\alpha=0.01$ (Middle Panel); and (iii) plot of     $\lambda_O$  given by (\ref{eq:lambdaOTime}) (Right Panel).    The two  invertible branches of $\lambda_O(a)$ that yield $m_+$ and $m_-$ are plotted with different colour and line style.  
}\label{fig:cases}   
\end{figure}
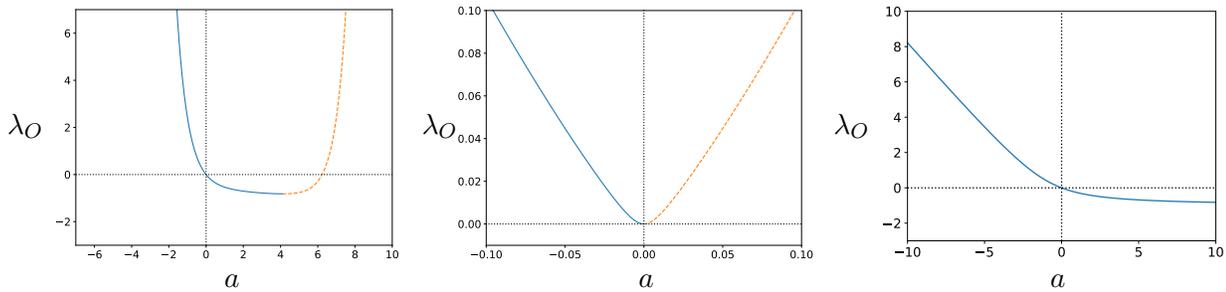

\section{Large deviation theory for  first-passage times of time-additive observables with nonzero average rates   ($\overline{o}> 0$)} \label{sec:moment} 
We derive  large deviation principles for $T$  in the limit  of large thresholds and for $\overline{o}> 0$.   In this case, the process $O_t$ is biased towards the positive threshold, and in the limit of large thresholds it holds that $p_+=1$.      
Hence, the events at the negative thresholds are suppressed.  We can distinguish two qualitatively different cases  depending on whether the events at the negative thresholds  are  {\it exponentially} or {\it super-exponentially} suppressed, and these two cases will be discussed separately.

\subsection{Main results}
The first-passage time $T$ satisfies a large deviation principle at  the positive and negative thresholds with speeds $\ell_+$ and $\ell_-$, respectively.  In other words, 
\begin{equation}
p_{T/\ell_+}(t|+) = \exp\left(-\ell_+  \: \mathcal{I}_+(t) [1+\mathcalligra{o}_{\ell_{\rm min}}(1)]\right)\label{eq:LDP1}
\end{equation}
and 
\begin{equation}
p_{T/\ell_-}(t|-) = \exp\left(-\ell_-  \: \mathcal{I}_-(t)[1+\mathcalligra{o}_{\ell_{\rm min}}(1)]\right), \label{eq:LDP2}
\end{equation}
where  $p_{T/\ell_+}(t|+)$ and $p_{T/\ell_-}(t|-)$ are the probability distributions of  $T/\ell_+$ and $T/\ell_-$ conditioned on events that terminate at the positive or negative threshold, respectively; where 
$\mathcal{I}_+(t)$ and $\mathcal{I}_-(t)$ are the corresponding rate functions~\cite{touchette2009large}; and where  $\mathcalligra{o}_{\:\ell_{\rm min}}(1)$ represents an arbitrary function that decays to zero  when $\ell_{\rm min} = {\rm min}\left\{\ell_-,\ell_+\right\}$ diverges.

 According to the G\"{a}rtner-Ellis theorem (see Theorem 2.3.6 in \cite{dembo2009large}), the rate functions $\mathcal{I}_+$ and $\mathcal{I}_-$  in Eqs.~(\ref{eq:LDP1}) and (\ref{eq:LDP2})  are the Fenchel-Legendre transforms of the scaled cumulant generating functions 
\begin{equation}
m_+(\mu) := \lim_{\ell_{\rm min}\rightarrow \infty} \frac{\ln g_+(\mu)}{\ell_+}
\end{equation}
and 
\begin{equation}
m_-(\mu) := \lim_{\ell_{\rm min}\rightarrow \infty} \frac{\ln g_-(\mu)}{\ell_-}  ,
\end{equation}  
respectively.

In this section, we use   the Perron martingale to derive expressions for  the scaled cumulant generating functions $m_+$ and $m_-$ in terms of the scaled cumulant generating function  $\lambda_O$, and for the splitting probability $p_-$ in the limit of large thresholds.     We distinguish two cases, 
depending on the number of solutions that the equation 
\begin{equation}
\lambda_O(a)  =  -\mu\label{eq:roots}
\end{equation} 
has,
namely: 
\begin{enumerate}
\item  {\it Equation~(\ref{eq:roots}) admits two solutions (e.g., as in the  Left Panel of Fig.~\ref{fig:cases})}:   the splitting probability $p_-$ decays exponentially fast as a function of $\ell_-$, viz., 
\begin{equation}
\lim_{\ell_{\rm min}\rightarrow \infty}  \frac{|\ln  p_-|}{\ell_-} = a^\ast >0 ,  \label{eq:pmin}
\end{equation}  
where $a^\ast$ is the nonzero root of the equation 
\begin{equation}
\lambda_O(a^\ast) = 0.  \label{eq:EADef}
\end{equation}  
Note that for fluctuating currents  $a^\ast$ is the effective affinity, and then it holds that $a^\ast \overline{j}\leq \dot{s}$ as shown in Refs.~\cite{neri2022universal,raghu2024effective}.

Furthermore, denoting the two solutions of 
 (\ref{eq:roots})  by   $-a_-(\mu)$ and $a_+(\mu)$,  with the convention that  $-a_-(\mu)<a_+(\mu)$, we  can express the scaled cumulant generating functions of $T$ at the positive and negative thresholds by 
\begin{equation}
m_+(\mu) =  -a_-(\mu)  \label{eq:mP}
\end{equation}  
and 
\begin{equation}
m_-(\mu) =  a^\ast -a_+(\mu), \label{eq:mM}
\end{equation}  
respectively.   
Hence, $m_+(-\mu)$ and $m_-(-\mu)$ are  the functional inverses corresponding with the two branches of $\lambda_O(a)$ (indicated with different line style in Fig.~\ref{fig:cases}), in agreement with  the results  for fluctuating currents  in Refs.~\cite{gingrich2017fundamenta, liu2024semi}.

\item  {{\it  Equation~(\ref{eq:roots}) admits one solution and $\lim_{a\rightarrow \infty}\lambda_O(a)$ is finite (e.g., as in  Right Panel of Fig.~\ref{fig:cases})}}: 
the splitting probability  at the negative threshold decays to zero faster than an exponential, i.e.,
\begin{equation}
\lim_{\ell_{\rm min}\rightarrow \infty}  \frac{|\ln  p_-|}{\ell_-} = \infty.  \label{eq:pmin2}
\end{equation}    
In addition, the  scaled cumulant generating function $m_+$ is given by 
\begin{equation}
m_+(\mu) =  a_+(\mu) , \label{eq:mP2}
\end{equation}  
where $a_+$ is the unique solution of (\ref{eq:roots}).  Hence, $m_+(-\mu)$ is the functional inverse of $\lambda_O(a)$, which is in agreement with the results in~Refs.~\cite{garrahan2017simple, liu2024semi}.    
 
\end{enumerate}

Fluctuating currents $J$ with a  nonzero average rate, $\overline{j}>0$, are examples of time-additive observables for which  (\ref{eq:roots}) admits two solutions, and thus the first scenario applies; an exception are fluctuating currents defined on a network that has unidirectional transitions so that $\mathbf{q}_{xy}>0$ and $\mathbf{q}_{yx}=0$, in which case it is possible that (\ref{eq:roots}) admits one solution.   The Left Panel of Fig.~\ref{fig:cases}     plots $\lambda_O(a)$ for a fluctuating current in the   random walker process  on a two-dimensional lattice  that we study  in Sec.~\ref{sec:randomwalker}.    

Examples for the second scenario are time-additive observables for which $O_t\geq 0$,  e.g.,  the time that $X$ has spent on one or more  states of the set $\mathcal{X}$ as in the Right Panel of  Fig.~\ref{fig:cases}, or time-additive observables that count the number of jumps  along one or more edges of  $\mathcal{E}$.   

There is a third scenario that we treat  separately in Sec.~\ref{sec:LDPMean}, namely when $\overline{o}=0$, as illustrated in the middle panel of Fig.~\ref{fig:cases}.      In this case the large deviation principles given by  Eqs.~(\ref{eq:LDP1}) and (\ref{eq:LDP2}) do not apply.   That the case $\overline{o}=0$ has to be treated differently can be understood from the Eqs.~(\ref{eq:pmin}), (\ref{eq:mP}) and (\ref{eq:mM}).     Indeed, when $\overline{o}=0$, then $a^\ast=0$, and hence according  to  (\ref{eq:pmin}) the splitting probability decays subexponentially as a function of $\ell_-$.     Moreover, for $\overline{o}=0$ it holds that $\lim_{\ell_+\rightarrow \infty}\langle T\rangle/\ell_+ = \left.\partial_\mu m_+(\mu)\right|_{\mu=0} = \infty$, implying that $\langle T\rangle$ scales super-linearly as a function of  $\ell_+$.

In what remains of    this Section, we   derive the Eqs.~(\ref{eq:mP}), (\ref{eq:mM}) and (\ref{eq:mP2}) with martingale theory.

\subsection{Doob's optional stopping theorem}
We apply
Doob's optional stopping theorem~\cite{roldan2022martingales}
\begin{equation}
\langle M_T\rangle = \langle M_0\rangle \label{eq:doobTh}
\end{equation}  
to the Perron martingale $M_t$ given by (\ref{eq:Mart}).   Doob's optional stopping theorem states that (\ref{eq:doobTh}) holds for the martingale $M_t$ if (i) the first-passage time $T$ is with probability one finite; and (ii)  $|M_t|$ is bounded for all values $t\in [0,T]$. Both conditions are satisfied as long as the thresholds $\ell_-$ and $\ell_+$ are finite and $\lambda_O(a)\geq 0$.   In this paper, we  assume that   (\ref{eq:doobTh}) also applies   when $\lambda_O(a)<0$, but we leave the  proof of this claim  open for  future work.    
Note that Doob's optional stopping Eq.~(\ref{eq:doobTh}) is central  to this work, as   all    calculations in this Paper boil down to using Eq.~(\ref{eq:doobTh}) on the  martingales (\ref{eq:Mart2}).

Substituting the  Perron martingale (\ref{eq:Mart}) into (\ref{eq:doobTh}) yields the equations 
\begin{equation}
 p_+ \langle \phi_a(X_T) e^{-aO_T-\lambda_O(a)T}\rangle_+  + p_- \langle \phi_a(X_T) e^{-aO_T-\lambda_O(a)T}\rangle_- = \langle \phi_a(X_0) \rangle  \label{eq:doob}
\end{equation}  
for all $a\in \mathbb{R}$.      To derive first-passage quantities from the  Eqs.~(\ref{eq:doob}), we need to deal with the  correlations between the random variables    $X_T$, $O_T$ and $T$ in the expected values.   Analytically tractable cases often correspond with scenarios when the random variables  $X_T$, $O_T$ and $T$  decorrelate.   In the present section,  we consider such a scenario, namely when both thresholds $\ell_-$ and $\ell_+$ are large.  

Since $O_t$ is a time-additive observable with increments that are independent of $t$, it holds that 
\begin{equation}
O_T = \ell_+ (1+\mathcalligra{o}_{\:\ell_{+}}(1)) \quad {\rm and} \quad O_T = -\ell_- (1+\mathcalligra{o}_{\:\ell_{-}}(1)), \label{eq:OAsymp}
\end{equation}
where the two expressions hold within the conditional averages $\langle \cdot\rangle_+$ and $\langle \cdot\rangle_-$, respectively. 
Substituting (\ref{eq:OAsymp})  in (\ref{eq:doob}) yields 
 \begin{equation}
p_+ e^{-a\ell_+(1+\mathcalligra{o}_{\:\ell_{+}}(1))}\Big\langle e^{-\lambda_O(a)T}\Big\rangle_+ + p_-  e^{a\ell_-(1+\mathcalligra{o}_{\:\ell_{-}}(1))}\Big\langle e^{-\lambda_O(a)T}\Big\rangle_- = 1, \label{eq:doobApplied}
\end{equation}
where we have  absorbed  $\ln( \phi_a(x)/\langle\phi_a(X_0)\rangle)$  in  $\mathcalligra{o}_{\ell_+}(1)$ and $\mathcalligra{o}_{\ell_-}(1)$,   as  for the Perron martingale $\phi_a(x)$ is positive and bounded.  Notice  that this latter assumption is not necessarily valid when the set $\mathcal{X}$ has infinite cardinality.  
Next we discuss   two qualitatively different cases, depending on whether  the equation   $\lambda_O(a)=0$ has one or two roots.

\subsection{Scaled cumulant generating functions of $T$ when events at the negative threshold  are exponentially suppressed}     \label{subsec:exp}

We derive  the Eqs.~(\ref{eq:pmin}), (\ref{eq:mP}) and (\ref{eq:mM}) when $\lambda_O(a)=-\mu$ has two nonzero solutions.

 Setting  $a=0$ and $a=a^\ast$ in Eq.~(\ref{eq:doobApplied}) yields  
\begin{equation}
p_- + p_+ = 1,
\end{equation}
and   
\begin{equation}
p_+ e^{-a^\ast \ell_+(1+\mathcalligra{o}_{\:\ell_{+}}(1))} + p_-  e^{a^\ast \ell_-(1+\mathcalligra{o}_{\:\ell_{-}}(1))} = 1, \label{eq:doobAppliedv2}
\end{equation}
respectively.  
Solving this set of two linear equations for $p_-$,  taking the  limit $\ell_{\rm min} = {\rm min}\left\{\ell_+,\ell_-\right\}\rightarrow \infty$, and using that $a^\ast>0$ for $\overline{o}>0$, we obtain the  Eq.~(\ref{eq:pmin}) for the exponential decay constant of the splitting probability.

Next, we derive expressions for the  scaled cumulant generating functions $m_+$ and $m_-$ of $T$.  Setting  $\lambda_O(a)=-\mu$  in (\ref{eq:doobApplied}),  and using that the former equation has two solutions, $a=a_+(\mu)$ and $a=-a_-(\mu)$, we obtain the equations
\begin{equation}
p_+ e^{-a_+(\mu)\ell_+(1+\mathcalligra{o}_{\ell_{+}}(1))}g_+(\mu) + p_-  e^{a_+(\mu)\ell_-(1+\mathcalligra{o}_{\ell_{-}}(1))}g_-(\mu) = 1, \label{eq:gPIntermed}
\end{equation}
and  
\begin{equation}
p_+ e^{a_-(\mu)\ell_+(1+\mathcalligra{o}_{\ell_{+}}(1))}g_+(\mu) + p_-  e^{-a_-(\mu)\ell_-(1+\mathcalligra{o}_{\ell_{-}}(1))}g_-(\mu) = 1.  \label{eq:gMIntermed}
\end{equation}   
Using the expressions $p_+ = 1 + \mathcal{O}\left(\exp(-a^\ast \ell_-)\right)$ and $p_- = \exp(-a^\ast \ell_- [1+\mathcalligra{o}_{\ell_{-}}(1)])$ for the splitting probabilities, we get 
\begin{equation}
e^{-a_+(\mu)\ell_+(1+\mathcalligra{o}_{\ell_{+}}(1))}g_+(\mu) + e^{(a_+(\mu)-a^\ast)\ell_-(1+\mathcalligra{o}_{\ell_{-}}(1))}g_-(\mu) = 1, \label{eq:first}
\end{equation}
and 
\begin{equation}
e^{a_-(\mu)\ell_+(1+\mathcalligra{o}_{\ell_{+}}(1))}g_+(\mu) + e^{-(a_-(\mu)+a^\ast)\ell_-(1+\mathcalligra{o}_{\ell_{-}}(1))}g_-(\mu) = 1.   \label{eq:second}
\end{equation}
Solving Eqs.~(\ref{eq:first}) and (\ref{eq:second})  for $g_-$ and $g_+$,  we obtain 
 \begin{equation}
g_+(\mu) =\frac{e^{a_+(\mu)\ell_-}-e^{-a_-(\mu)\ell_-}}{e^{a_+(\mu)\ell_- + a_-(\mu)\ell_+}-e^{-a_-(\mu) \ell_- - a_+(\mu)\ell_+}}\label{eq:gP}
\end{equation}
and 
\begin{equation}
g_-(\mu) =e^{a^\ast \ell_-} \frac{e^{a_-(\mu)\ell_+}-e^{-a_+(\mu)\ell_+}}{e^{a_+(\mu)\ell_- + a_-(\mu)\ell_+}-e^{-a_-(\mu) \ell_- - a_+(\mu)\ell_+}}, \label{eq:gM}
\end{equation} 
where we have  omitted the $\mathcalligra{o}_{\ell_{\pm}}(1)$  terms in the exponents.   Using that in our notation $a_+>-a_-$, we find 
\begin{equation}
m_+(\mu) = \lim_{\ell_{\rm min}\rightarrow\infty}\frac{\ln g_+(\mu)}{\ell_+} =  -a_-(\mu)  \label{eq:mPx}
\end{equation}
and 
\begin{equation}
m_-(\mu) = \lim_{\ell_-\rightarrow\infty}\frac{\ln g_-(\mu)}{\ell_-} =  a^\ast -a_+(\mu) , \label{eq:mMx}
\end{equation}  
which are the Eqs.~(\ref{eq:mP})  and (\ref{eq:mM}) that we were meant to derive.

\subsection{Cumulant generating functions of $T$ when events at the negative threshold  are super-exponentially suppressed}   
We derive the Eqs.~(\ref{eq:pmin2}) and (\ref{eq:mP2}) for the case when  Eq.~(\ref{eq:roots}) admits one solution, 
$\lambda_O(a_+)=-\mu$ and with the assumption that $\lim_{a\rightarrow \infty}\lambda_O(a)$ is finite. 

As $\lim_{a\rightarrow \infty}\lambda_O(a)\in \mathbb{R}$, the splitting probability $p_-$ decays super-exponentially, and therefore Eq.~(\ref{eq:pmin2}) holds.  

Since $p_-$ decays super-exponentially as a function of $\ell_-$,  the second term in Eq.~(\ref{eq:doob}) vanishes when $\ell_-$ is large enough, yielding 
\begin{equation}
 p_+ \langle \phi_a(X_T) e^{-aO_T-\lambda_O(a)T}\rangle_+  = \langle \phi_a(X_0) \rangle.  \label{eq:doob2}
\end{equation}  
Using  that $p_+ = 1 + \mathcalligra{o}_{\ell_{\rm min}}(1) $,  $\lambda_O(a_+)=-\mu$, $O_T=\ell_+(1+\mathcalligra{o}_{\ell_{\rm min}}(1))$, and  that $\phi_a(x)$ is positive, bounded, and independent of the thresholds,  we find  that (\ref{eq:doob2}) simplifies into
\begin{equation}
 e^{-a_+(\mu)\ell_+(1+\mathcalligra{o}_{\:\ell_{\rm min}}(1))}g_+(\mu)  = 1,
\end{equation}
and thus  we recover  the Eq.~(\ref{eq:mP2})  that we were meant to derive.

\section{Thermodynamic bounds on the cumulant generating functions of fluctuating currents}\label{sec:thermoBound}

For fluctuating currents, i.e., $O_t=J_t$,  the scaled cumulant generating function $m_+$ of  $T$ is lower bounded by~\cite{gingrich2016dissipation} 
\begin{equation}
m_+(\mu) \geq \frac{\dot{s}}{2\overline{j}} \left(1-\sqrt{1-4\mu/\dot{s} }\right),\label{eq:bound1x}
\end{equation} 
where $\dot{s}$ is the rate of dissipation given by Eq.~(\ref{eq:sdotDef}).  
On the right hand side of (\ref{eq:bound1x}) we recognise the scaled cumulant generating function of the inverse Gaussian distribution 
\begin{equation}
\sqrt{\frac{\ell^2_+\dot{s} }{4\pi t^3 \overline{j}^2}} \exp\left(-\frac{(t-\ell_+/\overline{j})^2 \dot{s}}{4t}\right), \label{eq:IG}
\end{equation}
with a shape parameter that is determined by the rate of dissipation $\dot{s}$,
and therefore (\ref{eq:bound1x}) yields the thermodynamic uncertainty relation given by Eqs.~(\ref{eq:thermo1}) and (\ref{eq:thermo2}).   Here, we derive a similar bound for $m_-$, the scaled cumulant generating function of $T$ at the negative threshold.

\subsection{Main results}
For fluctuating currents, we show that
\begin{equation}
m_-(\mu) \geq  
 \frac{\dot{s}}{2\overline{j}} \left(1-\sqrt{1-4\mu/\dot{s} }\right) +  a^\ast-\frac{\dot{s}}{\overline{j}} .\label{eq:bound2x}
\end{equation}   
The right-hand side of Eq.~(\ref{eq:bound2x}) is the scaled cumulant generating function of an inverse Gaussian distribution, as given by (\ref{eq:IG}) but $\ell_+$ substituted for $\ell_-$, plus an additional term given by $a^\ast - \dot{s}/\overline{j}$.   
The additional term is nonpositive, as~\cite{raghu2024effective},  
\begin{equation}
a^\ast \overline{j} \leq \dot{s}, \label{eq:EABound}
\end{equation}
and thus, the inequality (\ref{eq:bound2x}) 
 does  in general not imply a thermodynamic uncertainty relation for first-passage times at  negative thresholds.    

Nevertheless, we  show that the thermodynamic uncertainty relation 
for first-passage times at  negative thresholds, 
 \begin{equation}
 \dot{s} \geq 2 \frac{\langle T \rangle_-}{\langle T^2\rangle_- - \langle T \rangle^2_- } (1+\mathcalligra{o}_{\ell_{\rm min}}(1)), \label{eq:TM}
 \end{equation} 
holds for two sets of fluctuating currents.   The first set contains fluctuating currents that satisfy the Gallavotti-Cohen-like fluctuation symmetry~\cite{lebowitz1999gallavotti} 
\begin{equation}
\lambda_J(a) = \lambda_J(a^\ast-a). \label{eq:lebGol}
\end{equation}
Indeed, currents for which (\ref{eq:lebGol}) holds satisfy 
 the    
first-passage-time  symmetry relation~\cite{saito2016waiting,  decision1, gingrich2017fundamenta,neri2017statistics, roldan2022martingales} 
     \begin{equation}
     m_-(\mu) = m_+(\mu),  \label{eq:symm}
     \end{equation}
     and hence the  right-hand side of (\ref{eq:TM}) equals the corresponding ratio at the positive threshold.     The second set of currents for which the thermodynamic uncertainty relation applies at  negative thresholds are optimal currents for which 
      the equality in   (\ref{eq:EABound}) is attained.   The fluctuating entropy production  $S_t$, and  currents that are in the same cycle equivalence class as $S_t$~\cite{raghu2024effective}, are examples of currents that  are both optimal and satisfy  the Gallavotti-Cohen symmetry relation.     However,  in general, currents that satisfy the Gallavotti-Cohen   symmetry relation are not guaranteed to be optimal, and hence optimal currents are distinct from "symmetrical" currents.

\subsection{Derivation of the bound  (\ref{eq:bound2x})}
The scaled cumulant generating function of $J$ is bounded from below by~\cite{pietzonka} 
 \begin{equation}
 \lambda_J(a) \geq a \overline{j}\left(-1+a\frac{\overline{j}}{\dot{s}}\right).\label{eq:bound2xx}
 \end{equation}  
 The inequality~(\ref{eq:bound2xx}) can be derived with  the theory of level 2.5 deviations~\cite{ gingrich2016dissipation}.
 For fluctuating currents, Eq.~(\ref{eq:roots})  has two roots $-a_-(\mu)<a_+(\mu)$, and the inequality (\ref{eq:bound2xx}) implies that 
 \begin{equation}
a_+(\mu) \leq  \frac{\dot{s} + \sqrt{\dot{s}^2-4\mu \dot{s}}}{2\overline{j}} \label{eq:aPIneq}
\end{equation}
and 
\begin{equation}
a_-(\mu) \leq \frac{-\dot{s} +\sqrt{\dot{s}^2-4\mu \dot{s}}}{2\overline{j}} .  \label{eq:aMIneq}
\end{equation} 
 Using  the inequality (\ref{eq:aMIneq}) in Eq.~(\ref{eq:mP})  yields the inequality (\ref{eq:bound1x}), and substituting  (\ref{eq:aPIneq}) in (\ref{eq:mM}) we find the inequality (\ref{eq:bound2x}) that we were meant to derive.

\subsection{Derivation of the thermodynamic uncertainty relation at  negative thresholds}\label{sec:symm}
First, we derive the thermodynamic uncertainty relation (\ref{eq:TM}) for currents that satisfy the Gallavotti-Cohen-like fluctuation symmetry (\ref{eq:lebGol}).   

Due to (\ref{eq:lebGol}), the two roots $-a_-(\mu)$ and $a_+(\mu)$ of the equation $\lambda_J(a) = - \mu$ are related by 
\begin{equation}
-a_-(\mu) = a^\ast -a_+(\mu) .   \label{eq:tempEq}
\end{equation}
Using Eqs.~(\ref{eq:mPx}) and (\ref{eq:mMx}) in (\ref{eq:tempEq}), we obtain the first-passage-time symmetry relation Eq.~(\ref{eq:symm}).  

The thermodynamic uncertainty relation (\ref{eq:TM}) follows from the symmetry relation (\ref{eq:symm}) and the thermodynamic uncertainty relation at the positive threshold~\cite{gingrich2017fundamenta},
 \begin{equation}
 \dot{s} \geq 2 \frac{\langle T \rangle_+}{\langle T^2\rangle_+ - \langle T \rangle^2_+ } (1+\mathcalligra{o}_{\ell_{\rm min}}(1)). \label{eq:TP}
 \end{equation}  

 For optimal currents with $\dot{s}=a^\ast\overline{j}$, the inequality (\ref{eq:bound2x})   readily implies the   thermodynamic uncertainty relation (\ref{eq:TM}) at the negative threshold~\cite{gingrich2017fundamenta}.   However, in this case the right-hand side of (\ref{eq:TM}) is not guaranteed to be the same as the right-hand side of (\ref{eq:TP}).

\section{Scaled cumulants of the first-passage times of time-additive observables  with nonzero average rates ($\overline{o}>0$)      } \label{sec:moment2}

We determine   the scaled cumulants of $T$ in the limit of large thresholds, $\ell_-\gg 1$ and $\ell_+\gg1$,  for time-additive observables $O_t$ that have a nonzero average rate of change, $\overline{o}>0$.   

\subsection{Main results}
At the positive threshold, the first two cumulants of $T$  are  determined by $\overline{o}$ and the diffusivity coefficient $\sigma^2_O$, viz.,
\begin{equation}
\lim_{\ell_+\rightarrow \infty}\frac{\langle T\rangle_+}{\ell_+} = \frac{1}{\overline{o}} \label{eq:omean}
\end{equation}
and 
\begin{equation}
\lim_{\ell_+\rightarrow \infty}\frac{\langle T^2\rangle_+-\langle T\rangle^2_+}{\ell_+}  = \frac{\sigma^2_O}{\overline{o}^3}.  \label{eq:ovariance}
\end{equation}

On the other hand the cumulants of $T$ at the negative threshold  are determined by a Markov process that has the transition rate matrix
\begin{equation}
\mathbf{q}^\ast_{xy} := \frac{1}{\phi_{a^\ast}(x)}\tilde{\mathbf{q}}_{xy}(a^\ast) \phi_{a^\ast}(y), \label{eq:qDual}
\end{equation}
where $\tilde{\mathbf{q}}(a)$ is the tilted matrix defined in Eq.~(\ref{eq:qtilde}), where $a^\ast$ is the nonzero root of the scaled cumulant generating function $\lambda_O(a)$ [see Eq.~(\ref{eq:EADef})], and where $\phi_a(x)$ is the right eigenvector associated with the Perron root of  $\tilde{\mathbf{q}}(a)$.   We call    the Markov jump process defined by $\mathbf{q}^\ast$   the {\it dual} process associated with  $O_t$.  Importantly, the cumulants of $T$ at the negative threshold are determined by the cumulants of $O_t$ in the dual process, i.e.,    
\begin{equation}
\lim_{\ell_{\rm min}\rightarrow \infty}\frac{\langle T\rangle_-}{ \ell_-} = \frac{1}{|\overline{o}^\ast|  }\label{eq:TAvoAst}
\end{equation}
and 
\begin{equation}
\lim_{\ell_{\rm min}\rightarrow \infty}\frac{\langle T^2\rangle_--\langle T\rangle^2_-}{\ell_-}  = \frac{(\sigma^\ast_O)^2}{|\overline{o}^\ast|^3} , \label{eq:TSigmaJEA}
\end{equation}
where the $\ast$ indicates that the averages are with respect to the dual  process; notice that we use  absolute values $|\overline{o}^\ast|$ as $\overline{o}^\ast<0$.  The average rate  of $O_t$ in the dual process is given by 
\begin{equation}
\overline{o}^\ast  :=  \lim_{t\rightarrow \infty} \frac{\langle O_t\rangle_\ast}{t} =  \sum_{x\in \mathcal{X}}c_x \: p^\ast_{\rm ss}(x)+ \sum_{(x,y)\in \mathcal{E}} c_{xy} \:p^\ast_{\rm ss}(x)\mathbf{q}^\ast_{xy} ,\label{eq:TMinEA}
\end{equation}
where $p^\ast_{\rm ss}$ is the stationary distribution of the Markov process with rate matrix $\mathbf{q}^\ast$.  Analogously,    $(\sigma^\ast_O)^2$ is the diffusivity constant of $O_t$ when the statistics of  $X_t$ are drawn from   the dual process.    Hence,   $(\sigma^\ast_O)^2$   takes the form of Eq.~(\ref{eq:diffusivity}), albeit with $\langle \cdot \rangle$ replaced by $\langle \cdot \rangle_\ast$.    Note that to derive the formulae (\ref{eq:TAvoAst}) and (\ref{eq:TSigmaJEA}) we use that Eq.~(\ref{eq:roots}) admits two solutions, and hence the Eqs.~(\ref{eq:TAvoAst}) and (\ref{eq:TSigmaJEA}) hold for observables $O$ whose splitting probability $p_-$  is an exponentially decaying function of $\ell_-$.

\subsection{Cumulants of $T$ at the positive threshold} 
We derive the Eqs.~(\ref{eq:omean}) and (\ref{eq:ovariance}) that relate the cumulants of $O$ with the cumulants of $T$ conditioned on  first arrival at the positive threshold $\ell_+$.    According to   (\ref{eq:mP}), $-a_-(\mu)$ is  the scaled cumulant generating function $m_+$ that determines the cumulants  
of $T$ for trajectories that terminate at the positive threshold,  and thus
\begin{equation}
-a_-(\mu) =  \lim_{\ell_{\rm min}\rightarrow \infty}\frac{\langle T\rangle_+}{\ell_+} \mu  + \frac{1}{2}\lim_{\ell_{\rm min}\rightarrow \infty}\frac{\langle T^2\rangle_+ - \langle T\rangle^2_+ }{\ell_+} \mu^2 + \mathcal{O}(\mu^3)  .\label{eq:mPeq}
\end{equation}
Furthermore, by definition $-a_-$ solves  Eq.~(\ref{eq:roots}),  i.e., 
\begin{equation}
\lambda_O(-a_-(\mu)) = -\mu, \label{eq:lambdaPOSolve}
\end{equation} 
where  $\lambda_O$ is the cumulant generating function of $O$, and thus 
\begin{equation}
\lambda_O(a) = - \overline{o} \: a + \frac{\sigma^2_O}{2}  a^2+ \mathcal{O}(a^3). \label{eq:thirdOA}
\end{equation} 
Using (\ref{eq:thirdOA})  in (\ref{eq:lambdaPOSolve}), and 
solving for  $a_-$, we obtain 
\begin{equation}
-a_-(\mu) = \frac{\mu}{\overline{o}} + \frac{1}{2} \mu^2  \frac{\sigma^2_O}{\overline{o}^3} + \mathcal{O}(\mu^3).  
\label{eq:aMRes}
\end{equation}  
Identifying  the linear coefficients in Eqs.~(\ref{eq:mPeq}) and (\ref{eq:aMRes}) we obtain the equality Eq.~(\ref{eq:omean}),   
and identifying the corresponding quadratic coefficients results into the relation (\ref{eq:ovariance}).   
Analogously, we can obtain relations between the third order and higher order cumulants of $T$ and~$O$.

\subsection{Cumulants of $T$ at the negative threshold}\label{sec:dual}
We derive the Eqs.~(\ref{eq:TAvoAst}) and (\ref{eq:TSigmaJEA}) relating the cumulants of $T$ at the negative threshold with those of $O$ in the dual process.   For events terminating at the negative threshold, the scaled cumulant generating function $m_-$  equals $a^\ast -a_+(\mu)$, see Eq.~(\ref{eq:mM}), and thus 
\begin{equation}
a^\ast -a_+(\mu)  = \lim_{\ell_{\rm min}\rightarrow \infty}\frac{\langle T\rangle_-}{\ell_-} \mu  + \frac{1}{2}\lim_{\ell_{\rm min}\rightarrow \infty}\frac{\langle T^2\rangle_- - \langle T\rangle^2_- }{\ell_-} \mu^2 + \mathcal{O}(\mu^3)  .\label{eq:muMinSolve}
\end{equation} 
 In addition, as $a_+$ is a solution of Eq.~(\ref{eq:roots}) it holds that 
\begin{equation}
\lambda_O(a_+(\mu)) = -\mu.   \label{eq:lambdaOap12}
\end{equation}
Since  $a_+(0) = a^\ast$,  we consider the Taylor series of   $\lambda_O(a)$   at $a=a^\ast$, viz., 
\begin{equation}
\lambda_O(a) =  (a-a^\ast) \lambda'_O(a^\ast) + \frac{1}{2}(a-a^\ast)^2  \lambda''_O(a^\ast) + \mathcal{O}((a-a^\ast)^3) ,\label{lambdaOM}  
\end{equation}
where we have used that $\lambda_O(a^\ast)=0$ and the notation $\lambda'_O(a) = \partial_a \lambda_O(a)$ for the derivative of $\lambda_O(a)$ with respect to $a$.
Using (\ref{lambdaOM}) in (\ref{eq:lambdaOap12}) 
 and solving for  $a_+(\mu)$ yields 
\begin{equation}
a^\ast - a_+(\mu) = \frac{\mu}{\lambda'_O(a^\ast)} + \frac{\mu^2}{2}  \frac{\lambda''_O(a^\ast)}{[\lambda'_O(a^\ast)]^3} +\mathcal{O}(\mu^3).  \label{eq:MM}
\end{equation}
Identifying the linear coefficients in Eqs.~(\ref{eq:muMinSolve}) and (\ref{eq:MM}) yields 
\begin{equation}
\lim_{\ell_{\rm min}\rightarrow \infty}\frac{\langle T\rangle_-}{\ell_-}  = \frac{1}{\lambda'_O(a^\ast)} \label{eq:TMinL}
\end{equation}
and identifying    the quadratic coefficients we obtain 
\begin{equation}
\lim_{\ell_{\rm min}\rightarrow \infty}\frac{\langle T^2\rangle_- - \langle T \rangle^2_-}{\ell_-}  =   \frac{\lambda''_O(a^\ast)}{[\lambda'_O(a^\ast)]^3} . \label{eq:TMinL2} 
\end{equation}

The derivatives  of $\lambda_O(a)$ evaluated at $a=a^\ast$ determine the cumulants of $O$ in a dual process with rate matrix $\mathbf{q}^\ast$ [given by (\ref{eq:qDual})].  
Indeed, the tilted matrix of $\mathbf{q}^\ast$, which we denote by $\tilde{\mathbf{q}}^\ast(a)$, takes the form 
\begin{equation}
\tilde{\mathbf{q}}^\ast_{xy}(a) =\frac{1}{\phi_{a^\ast}(x)}\tilde{\mathbf{q}}_{xy}(a+a^\ast)\phi_{a^\ast}(y).
\end{equation}
Hence $\tilde{\mathbf{q}}^\ast(a)$ and $\tilde{\mathbf{q}}(a+a^\ast)$ have the same eigenvalues, as they are related by a similarity transformation.  Therefore,  the cumulant generating function  $\lambda^\ast_O$ of $O$ in the dual process (determined by $\mathbf{q}^\ast$) is related  to the cumulant generating function $\lambda_O$  of $O$ in the original process  (determined by $\mathbf{q}$)  through the equality
\begin{equation} 
\lambda^\ast_O(a) = \lambda_O(a+a^\ast).   \label{eq:lambdaRel}
\end{equation}
An interesting consequence of  Eq.~(\ref{eq:lambdaRel}) is that the effective affinity $a^{\ast \ast}$ of $O$ in the  dual process is given by  $a^{\ast \ast}= -a^\ast$, and thus the sign of $\overline{o}^\ast$ is negative.    It also follows from Eq.~(\ref{eq:lambdaRel}) that 
\begin{equation}
\overline{o}^\ast = -\partial_a \lambda^\ast_O(a)|_{a=0} = -  \partial_a \lambda_O(a)|_{a=a^\ast} 
\end{equation}
and 
\begin{equation}
(\sigma^\ast_O)^2 = \partial^2_a \lambda^\ast_O(a)|_{a=0} =  \partial^2_a \lambda_O(a)|_{a=a^\ast} . 
\end{equation}
Using these formulae, in Eqs.~(\ref{eq:TMinL}) and (\ref{eq:TMinL2}), we recover the Eqs.~(\ref{eq:TAvoAst}) and (\ref{eq:TSigmaJEA})  that we were meant to derive.

\section{Moment generating functions at finite thresholds ($\overline{o}>0$)} \label{sec:finite}
So far we have focused on first-passage problems at   large thresholds so that both $\ell_+\gg 1$ and $\ell_-\gg 1$.  In this limit, the  statistics of $T$ are fully determined by the cumulant generating function $\lambda_O(a)$ [as defined in Eq.~(\ref{def:lambdaO})].  However,  if the thresholds are finite then $\lambda_O(a)$ does not suffice to determine the splitting probabilities and the cumulant generating functions of $T$, which complicates the analysis at  finite thresholds. 
In this section,  we solve the first-passage problem of $O_t$ at finite thresholds for a  particular class of time-additive observables.   This analysis will reveal some of the differences between  first-passage problems at finite  and infinite thresholds.

Specifically, we consider observables $O_t$ for which  the pair $(X_T,O_T)$ is deterministic  when it is conditioned upon reaching either of the two thresholds, i.e.,
\begin{equation} (X_T,O_T) \in \left\{(y,\ell_+), (x,-\ell_-)\right\},  \label{eq:property}\end{equation} 
with $x$ and $y$ two fixed states in the set $\mathcal{X}$.   This implies that $X_T=y$ when $O_T=\ell_+$, and $X_T=x$ when $O_T=-\ell_-$.  

\subsection{Main results}
We consider time-additive observables of the form
\begin{equation}
O_t = \sum_{z\in \mathcal{X}: (z,y)\in \mathcal{E}}n_{z} N^{zy}_t - \sum_{z\in \mathcal{X}: (z,x)\in \mathcal{E}}m_{z} N^{zx}_t .   \label{eq:OtFinite}
\end{equation}  
with $n_z,m_z\in \left\{0,1\right\}$ coefficients that define $O_t$.   The first term in (\ref{eq:OtFinite}) describes   positive increments of $O_t$ corresponding with transitions  towards the $y$-state, while the second term  in (\ref{eq:OtFinite}) describes negative increments of $O_t$ corresponding with transitions towards the $x$-state.  Thus, $O_t$ satisfies  Eq.~(\ref{eq:property}).   
In addition, we consider observables $O_t$ for which  the equation $\lambda_O(a)=-\mu$ has two solutions, so that the events at the negative threshold are exponentially suppressed (see Sec.~\ref{sec:moment}).    Examples of  time-additive observables   that satisfy these  conditions are 
edge currents, 
\begin{equation}
J^{xy}_t = N^{xy}_t-N^{yx}_t,
\end{equation}  
which are obtained from   Eq.~(\ref{eq:OtFinite}) by setting   $n_z = \delta_{z,x}$ and $m_z = \delta_{z,y}$.   

If $\overline{o}>0$, then 
the splitting probability  of observables  $O_t$ of the form    (\ref{eq:OtFinite}) takes the form 
\begin{equation}
p_- = \frac{\langle \phi_{a^\ast}(X_0)\rangle - \phi_{a^\ast}(y) e^{-a^\ast \ell_+} }{\phi_{a^\ast}(x) e^{a^\ast \ell_-} - \phi_{a^\ast}(y) e^{-a^\ast \ell_+}}, \label{eq:pM2}
\end{equation}
where $a^\ast$ is the nonzero root of Eq.~(\ref{eq:EADef}).   The splitting probability at the positive threshold equals 
 $p_+=1-p_-$.

The generating functions $g_+$ and $g_-$, as defined in (\ref{eq:genFunc}), are given by
\begin{eqnarray}
g_+(\mu) 
= \frac{1}{p_+} \frac{e^{a_+\ell_-}  \phi_{a_+}(x)  \langle \phi_{-a_-}(X_0) \rangle - e^{-a_-\ell_-} \phi_{-a_-}(x) \langle \phi_{a_+}(X_0) \rangle}{e^{a_+\ell_-+a_-\ell_+} \phi_{-a_-}(y)\phi_{a_+}(x) - e^{-a_-\ell_--a_+\ell_+}\phi_{-a_-}(x)\phi_{a_+}(y) }  \label{eq:gP2}
\end{eqnarray}
and 
\begin{eqnarray}
 g_-(\mu) =  \frac{1}{p_-}  \frac{e^{a_-\ell_+}\phi_{-a_-}(y)\langle \phi_{a_+}(X_0)\rangle -e^{-a_+\ell_+}\phi_{a_+}(y)\langle \phi_{-a_-}(X_0)\rangle}{e^{a_+\ell_-+a_-\ell_+} \phi_{-a_-}(y)\phi_{a_+}(x) - e^{-a_-\ell_--a_+\ell_+}\phi_{-a_-}(x)\phi_{a_+}(y)}, \label{eq:gM2}
 \end{eqnarray}
respectively, 
where $-a_-(\mu)$ and $a_+(\mu)$ are the two roots of the equation $\lambda_O(a)= -\mu$ (for notational simplicity we omitted in (\ref{eq:gP2}) and (\ref{eq:gM2}) the functional dependency of $a_-$ and $a_+$ on  $\mu$).

Note that, excluding special cases,  the splitting probability $p_-$ and the generating functions $g_+$ and $g_-$  are determined by  both   the Perron root $\lambda_O(a)$ of the tilted matrix $\tilde{\mathbf{q}}(a)$ and   its corresponding  right eigenvector $\phi_a(x)$.    At large thresholds,  the dependency on the right eigenvector $\phi_a(x)$ becomes irrelevant, and we recover the generic results of Sec.~\ref{sec:moment}.  Hence, the fluctuations in $T$ at  finite thresholds, and the dependency on the initial condition $X_0$, are determined by the right eigenvector $\phi_a$.

\subsection{Doob's optional stopping theorem}
In the remainder part of this section,  we derive the Eqs.~(\ref{eq:pM2})-(\ref{eq:gP2}).   As before, the starting point is  the   Eq.~(\ref{eq:doob}), which is Doob's optional stopping theorem applied to the Perron martingale (\ref{eq:Mart}).    For observables of the form (\ref{eq:OtFinite})  the property (\ref{eq:property}) applies, and consequently  Eq.~(\ref{eq:doob})  simplifies into 
\begin{equation}
 p_+ \phi_a(y)  e^{-a\ell_+} \langle   e^{ -\lambda_O(a)T}\rangle_+  + p_- \phi_a(x) e^{a\ell_-}  \langle e^{-\lambda_O(a)T}\rangle_- = \langle \phi_a(X_0) \rangle   \label{eq:doobSP}
\end{equation}
for all $a\in \mathbb{R}$.

\subsection{Splitting probability}
Setting $a=0$ in Eq.~(\ref{eq:doobSP}), and using $\lambda_O(0)=0$ and $\phi_0(x) = 1$, yields
\begin{equation}
p_- + p_+ = 1.  \label{eq:PPfirst}
\end{equation}
By assumption, the equation $\lambda_O(a)=0$ has a nontrivial root $a=a^\ast$, and setting $a=a^\ast$ in Eq.~(\ref{eq:doobSP}) yields 
\begin{equation}
 p_+  \phi_{a^\ast}(y) e^{-a^\ast \ell_+} + p_- \phi_{a^\ast}(x) e^{a^\ast \ell_-} = \langle \phi_{a^\ast}(X_0)\rangle. \label{eq:PPsecond}
\end{equation}
Solving the Eqs.~(\ref{eq:PPfirst}) and  (\ref{eq:PPsecond})  for $p_-$ and $p_+$, we obtain the  expression (\ref{eq:pM2}) for $p_-$.

From Eq.~(\ref{eq:PPsecond}) we can study    the  limit where the   positive threshold is large, $\ell_+\gg 1$, while  the   negative threshold $\ell_-$ is kept finite.  In this limit, 
\begin{equation}
\lim_{\ell_{+}\rightarrow \infty } \frac{|\ln p_-|}{\ell_-} =  a^\ast + \frac{1}{\ell_-}\ln \frac{\phi_{a^\ast}(x)}{\langle \phi_{a^\ast}(X_0)\rangle}. \label{eq:pfiniteL}
\end{equation} 
Comparing Eq.~(\ref{eq:pfiniteL}) with (\ref{eq:pmin}), we conclude that    corrections to the splitting probability due to  finite negative thresholds are determined by the  right eigenvector $\phi_{a^\ast}$.  The finite threshold correction depends on the initial condition, and if  $X_0=x$, then the correction term vanishes so that 
\begin{equation}
p_- = \exp(-a^\ast \ell_-) + \mathcal{O}(\exp(-a^\ast \ell_+)). \label{eq:pfiniteL2}
\end{equation}

\subsection{Moment generating functions}
 Substitution of $a_+$ and $-a_-$, the two roots of  $\lambda_O(a)=-\mu$, into Eq.~(\ref{eq:doobSP}) results into the two equations
\begin{equation}
 p_+ \phi_{a_+}(y) e^{-a_+\ell_+}   g_+(\mu)  + p_- \phi_{a_+}(x)e^{a_+\ell_-}  g_-(\mu)= \langle \phi_{a_+}(X_0) \rangle    \label{eq:gp1}
\end{equation} 
and 
\begin{equation}
p_+ \phi_{-a_-}(y) e^{a_-\ell_+}   g_+(\mu) + p_- \phi_{-a_-}(x)e^{-a_-\ell_-}  g_-(\mu)= \langle \phi_{-a_-}(X_0) \rangle   , \label{eq:gp2}
\end{equation} 
respectively.
Solving the Eqs.~(\ref{eq:gp1}) and (\ref{eq:gp2}) for $g_-$ and $g_+$, we obtain the solutions~(\ref{eq:gP2}) and~(\ref{eq:gM2}).

With Eqs.~(\ref{eq:gP2})  and  (\ref{eq:gM2})   we can study the statistics of $T$ when one of the two thresholds  $\ell_-$ and $\ell_+$ is infinitely large, while the other one remains finite.

Let us first consider the statistics of $T$ at finite negative thresholds.   Taking the limit $\ell_+\gg1$ in Eq.~(\ref{eq:gM2}) and using that  $a_+>-a_-$, we obtain 
\begin{eqnarray}
 \lim_{\ell_+\rightarrow \infty} \frac{\ln g_-(\mu)}{\ell_-}  
&=&   a^\ast - a_+(\mu) + \frac{1}{\ell_-} \ln  \left(
\frac{\langle \phi_{a_+(\mu)}(X_0)\rangle }{\langle \phi_{a^\ast}(X_0)\rangle}\frac{\phi_{a^\ast}(x) }{\phi_{a_+(\mu)}(x)}\right) .  \label{eq:gMAsympt1}
\end{eqnarray}  
Comparing (\ref{eq:gMAsympt1}) with (\ref{eq:mPx}), we conclude that the  generating function of $T$ at finite negative thresholds contains a correction term that is determined by the right eigenvector $\phi_a(x)$ of the Perron root of $\tilde{\mathbf{q}}(a)$ and depends on the initial state $X_0$.  
If we set $X_0=x$, then the correction term vanishes and we obtain the simpler formula
\begin{equation}
g_-(\mu)  =  e^{(a^\ast-a_+(\mu))\ell_-} \left(1 + \mathcal{O}(e^{-(a_-+a_+) \ell_+}) \right)  \label{eq:gMAsympt2}
\end{equation}  
that is independent of the right eigenvector $\phi_a$.   

Second we consider the statistics at a finite positive threshold.    Taking the limit $\ell_-\gg 1$ in Eq.~(\ref{eq:gP2})  and using that $p_+=1$ in this limit, we find 
\begin{equation}
\lim_{\ell_-\rightarrow \infty}\frac{\ln g_+(\mu)}{\ell_+} =-a_-(\mu) + \frac{1}{\ell_+}\ln  \left(\frac{\langle \phi_{-a_-(\mu)}(X_0)\rangle }{\phi_{-a_{-}(\mu)}(y)} \right) . \label{eq:gPfinite}
\end{equation}
As before, (\ref{eq:gPfinite}) equals the asymptotic result 
(\ref{eq:mP}) plus a correction term that depends on the initial state and vanishes in the limit of large thresholds.     If we furthermore constrain the initial condition to  $X_0=y$, then 
\begin{equation}
g_+(\mu) =    e^{-a_-(\mu)\ell_+}  \left(1+ \mathcal{O}(e^{-(a_-+a_+) \ell_-})\right). \label{eq:gP22}
\end{equation}
 Hence, the thermodynamic bound (\ref{eq:bound1x})  applies at finite $\ell_+$-thresholds in the specific case  of observables $O_t$ of the form (\ref{eq:OtFinite}) and for the initial condition $X_0=y$.  Consequently,  in this case also the thermodynamic uncertainty relation for first-passage times  applies at finite thresholds, i.e., 
 \begin{equation}
 \dot{s} \geq 2\frac{\langle T\rangle_+}{\langle T^2\rangle_+-\langle T\rangle^2_+} (1+\mathcalligra{o}_{\ell_-}(1)). \label{eq:FTTUR}
 \end{equation}

\section{Splitting probabilities and mean first passage times for  time-additive observables with zero average rates   ($\overline{o} = 0$)} 
 \label{sec:LDPMean}

We consider the case     $\overline{o}=0$, for which the observable $O_t$ behaves diffusively  as a function of $t$.   Similarly, we expect  that in this case diffusive properties appear in the first-passage properties of $O$.

If  $\overline{o}=0$, then the large deviation principles for $T$, as expressed by the Eqs.~(\ref{eq:LDP1}) and (\ref{eq:LDP2}), do not apply.       Indeed, the large deviation principle for $T$ follows from the Eqs.~(\ref{eq:mP}) and  (\ref{eq:mM}) and the  G\"{a}rtner-Ellis theorem (see Theorem 2.3.6 in \cite{dembo2009large}).   The  G\"{a}rtner-Ellis theorem  requires that the functions $m_+$ and $m_-$ exist and that in addition the  interiors of the sets $\mathcal{D}_+ = \left\{\mu\in \mathbb{R}:m_+(\mu)<\infty\right\}$ and  $\mathcal{D}_- = \left\{\mu\in \mathbb{R}:m_-(\mu)<\infty\right\}$ contain the  origin.   However, for $\overline{o}=0$,  the origin is a boundary point of the sets  $\mathcal{D}_+$
 and $\mathcal{D}_-$, and hence the origin does not belong to their interiors.   Consequently, the G\"{a}rtner-Ellis theorem  does not apply when $\overline{o}=0$ and we cannot conclude from the   Eqs.~(\ref{eq:mP}) and  (\ref{eq:mM}) that $T$ satisfies a large deviation principle of the form (\ref{eq:LDP1}) and (\ref{eq:LDP2}).  

Nevertheless, the processes $M_t$,  given by Eq.~(\ref{eq:Mart2}),  are  martingales when $\overline{o}=0$.   Hence,  also for  $\overline{o}=0$ we can use martingale theory to   derive explicit expressions for  splitting probabilities and  mean first-passage times, and this is the problem we address in this section.

 \subsection{Main results}

For time-additive observables $O_t$ that have zero average rate, $\overline{o}=0$, and a  nonzero diffusivity constant, $\sigma^2_O>0$, the splitting probability takes the expression
\begin{equation}
p_- = \frac{\langle O_T\rangle_+ \phi_0  +  \langle \phi'_0(X_0)\rangle  - \langle \phi'_0(X_T)\rangle_+  }{(\langle O_T\rangle_+ - \langle O_T\rangle_-) \phi_0 + \langle \phi'_0(X_T)\rangle_- -\langle \phi'_0(X_T)\rangle_+  }   \label{eq:pMZero}
\end{equation}  
and a corresponding equation holds for  $p_+ = 1-p_-$.  Here, $\phi'_a(x) = \partial_a \phi_a(x)$ and $\phi'_0(x) = \left.\partial_a \phi_a(x)\right|_{a=0}$.  Under the same conditions, the  mean first-passage time is given by
\begin{eqnarray}
 \langle T\rangle   &=&  \frac{p_+}{\sigma^2_O}  \left( \langle O^2_T \rangle_+ - 2 \frac{\langle O_T \phi'_0(X_T) \rangle_+}{\phi_0}  + \frac{\langle \phi''_0(X_T)\rangle_+ }{\phi_0} \right)  
\nonumber\\ 
&& 
+\frac{p_-}{\sigma^2_O }  \left( \langle O^2_T \rangle_- - 2 \frac{\langle O_T \phi'_0(X_T) \rangle_-}{\phi_0} +  \frac{\langle \phi''_0(X_T)\rangle_-}{\phi_0}\right) 
-\frac{1}{\sigma^2_O}\frac{\langle   \phi''_0(X_0)\rangle}{\phi_0}  .  \nonumber\\ \label{eq:meanTZero}
\end{eqnarray}

The diffusive behaviour of the observable $O_t$ becomes apparent when taking the limit of  large thresholds, $\ell_-,\ell_+\gg 1$.  In this case,   the splitting probabilities  are 
\begin{equation}
p_- =  \frac{\ell_+}{\ell_-+\ell_+}   + \mathcal{O}\left(1/\ell_{\rm min}\right) \quad {\rm and } \quad p_+ =   \frac{\ell_-}{\ell_-+\ell_+}  + \mathcal{O}\left(1/\ell_{\rm min}\right) , \label{eq:pMZeroO}
\end{equation} 
and for the mean first-passage time we find   that
\begin{equation}
\langle T\rangle =  \frac{p_+ \ell^2_+ + p_- \ell^2_-}{\sigma^2_O} + \mathcal{O}\left(\ell_+,\ell_-\right)  =  \frac{\ell_+\ell_-}{\sigma^2_O} + \mathcal{O}\left(\ell_+,\ell_-\right) , \label{eq:meanTAsymptoZero}
\end{equation}
which we recognise as the splitting probabilities and mean first-passage time of a  Brownian motion with diffusivity $\sigma^2_O$.  

Comparing the asymptotic formulae (\ref{eq:pMZeroO}) and (\ref{eq:meanTAsymptoZero}) with  those at finite thresholds, (\ref{eq:pMZero}) and (\ref{eq:meanTZero}), respectively,  we observe that correction terms due to  finite thresholds depend on the   statistics of the initial state $X_0$ and the properties of the right eigenvector $\phi_a(x)$ near $a\approx 0$.   Note that the Eqs.~(\ref{eq:pMZeroO})  and  (\ref{eq:meanTAsymptoZero}) are consistent with the Eqs.~(\ref{eq:pmin}) for $a^\ast=0$ and (\ref{eq:omean}) for $\overline{o}=0$, respectively.  

The Eqs.~(\ref{eq:pMZero})-(\ref{eq:meanTAsymptoZero}) require that $\sigma^2_O>0$, as otherwise there is a nonzero probability that $T=\infty$, and thus  Doob's optional stopping theorem, given by Eq.~(\ref{eq:doobTh}), does not apply; this is also evident from the divergence of (\ref{eq:meanTAsymptoZero}) in the limit $\sigma^2_O\rightarrow 0$.   
 An example of a time-additive observable for which $\sigma^2_O = 0$ is  the following: consider  a random walk process $X_t$ on a one-dimensional lattice with periodic boundary conditions, and consider a fluctuating current   with  the coefficients $c_{xy}$  defined to ensure that the current remains unchanged when $X_t$ makes a complete excursion through  the lattice.  In this case, the observable $O_t$ is bounded, i.e., there exists a constant $c$ so that $|O_t|<c$ for all values of $t$, and hence if the  thresholds are large enough they will not be reached.   
 
\subsection{Splitting probability: derivation}\label{subsec:subexp} 
To derive the Eq.~(\ref{eq:pMZero})  for the splitting probability $p_-$, we expand   the left and right-hand sides of  Doob's optional stopping equation (\ref{eq:doob})  in the variable  $a$, using that 
\begin{equation}
e^{-aO_T-\lambda_O(a)T} = 1 -a O_T + \mathcal{O}(a^2)   \label{eq:expansion1}
\end{equation}
and
\begin{equation}
\phi_a(x) = \phi_0 + a\phi'_0(x) + \mathcal{O}(a^2),  \label{eq:expansion2}
\end{equation}  
where  $\phi_0$ is a constant independent of $x$, and we have used that $\lambda'_O(0)=0$ for $\overline{o}=0$.   
Equating in (\ref{eq:doob}) the coefficients in zeroth order in $a$ yields the usual 
\begin{equation}
    p_-+p_+=1, \label{eq:firstx}
\end{equation}
and equating the   coefficients  in linear order in $a$ we get
\begin{equation}
p_-  (\langle \phi'_0(X_T)\rangle_-  -  \phi_0\langle O_T\rangle_-)  + p_+ (\langle \phi'_0(X_T)\rangle_+ - \phi_0\langle O_T\rangle_+) = \langle \phi'_0(X_0)\rangle. \label{eq:secondx} 
\end{equation}  
Solving the Eqs.~(\ref{eq:firstx}) and (\ref{eq:secondx}) for $p_-$ and $p_+$, yields  Eq.~(\ref{eq:pMZero}).

\subsection{Mean first-passage time: derivation}  
We  expand the left and right-hand side of Eq.~(\ref{eq:doob}) in $a$, albeit now we consider an expansion up to second order in $a$, viz., 
\begin{equation}
e^{-aO_T-\lambda_O(a)T} = 1 -a O_T -  \frac{a^2}{2} \left(\sigma^2_O T - O_T^2\right)   +  \mathcal{O}(a^3)   \label{eq:expansion1x}
\end{equation}
and
\begin{equation}
\phi_a(x) = \phi_0 + a\phi'_0(x) + \frac{a^2}{2} \phi''_0(x)  +  \mathcal{O}(a^3);\label{eq:expansion2x}
\end{equation} 
Notice that we have used 
$\lambda''_O(0) = \sigma^2_O$. 
Using these expressions in Eq.~(\ref{eq:doob}) and equating the coefficients on the left and right-hand side of the equality that appear in front of  $a^2$, we recover Eq.~(\ref{eq:meanTZero}). Note that this approach provides us with  an expression for $\langle T\rangle$,
but not for the conditional averages  $\langle T\rangle_+$ and  $\langle T\rangle_-$.
These follow instead from the observation that at large thresholds $O_t$ is well
approximated by a driftless diffusion with diffusivity $\sigma^2_O$, for which the
conditional mean exit times are known~\cite{redner2001guide}, viz.,
\begin{equation}
\langle T\rangle_+ = \frac{\ell^2_++2\ell_+\ell_-}{3\sigma^2_O} +\mathcal{O}(\ell_{\rm max})
\quad{\rm and} \quad
\langle T\rangle_- = \frac{\ell^2_-+2\ell_+\ell_-}{3\sigma^2_O} +\mathcal{O}(\ell_{\rm max}),
\label{eq:TCondZero}
\end{equation}
where $\ell_{\rm max} = {\rm max}\left\{\ell_-,\ell_+\right\}$.   For symmetric thresholds,
$\ell_-=\ell_+=\ell$, both expressions reduce to $\ell^2/\sigma^2_O$.   Note also that
together with the splitting probabilities (\ref{eq:pMZeroO}) they satisfy
$p_+\langle T\rangle_++p_-\langle T\rangle_-=\ell_+\ell_-/\sigma^2_O$, consistent with
Eq.~(\ref{eq:meanTAsymptoZero}).

\section{Random walker on a two-dimensional lattice:  a comparison of  the first-passage times statistics at  the positive and negative thresholds }\label{sec:randomwalker}  

\begin{figure}[h!]\centering
 \includegraphics[width=0.5\textwidth]{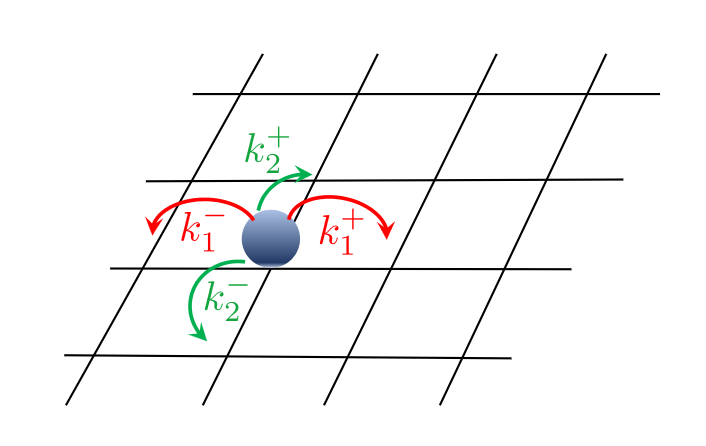}
{\includegraphics[width=0.4\textwidth]{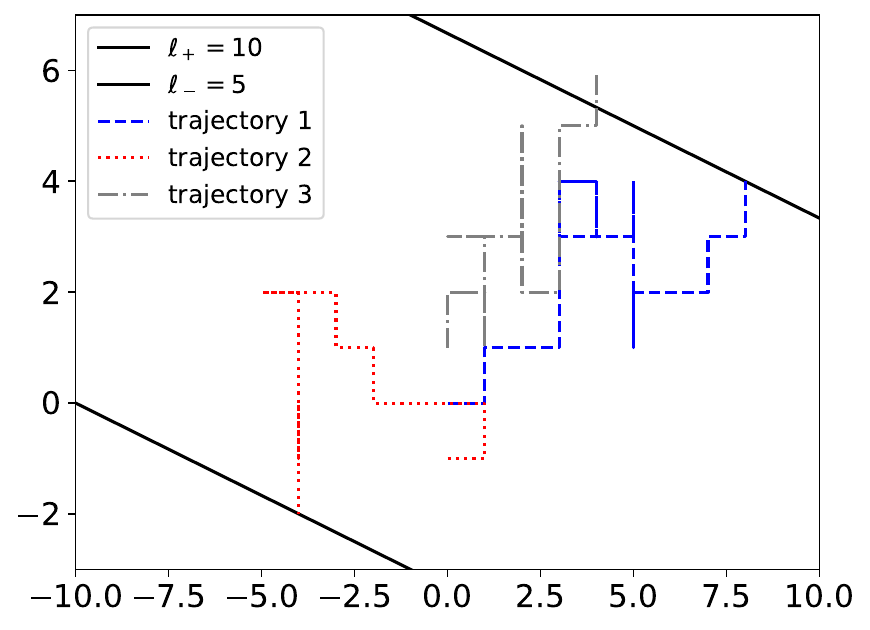}
\put(-90,-9){\small  $X^{(1)}$}
\put(-194,60){\small  $X^{(2)}$}}
\caption{ Escape problem of an active particle that leaves a strip in the two-dimensional plane of finite width.   Panel (a): illustration of the considered  random walk process on a two-dimensional lattice (Figure is taken from Ref.~\cite{neri2022estimating}).   Panel (b):   Three representative trajectories of $X_t$.   Parameters chosen are: the rates are given by  (\ref{eq:rate1})  and (\ref{eq:rate2}) with $\nu=1/2$ and $\rho = 1$,    the current  is given by (\ref{eq:Jexam}) with $\Delta = 1/2$,  and the threshold parameters are  set to  $\ell_-=5$ and $\ell_+=10$.   The length $n$ of the lattice  is taken to be large, $n\gg 1$.    }\label{fig:2DIllu}   
\end{figure}

We solve the first-passage problem of a biased random walker escaping from  a strip of finite width in the two-dimensional plane~\cite{neri2022estimating} (see the Left Panel  of Fig.~\ref{fig:2DIllu} for an illustration of the two-dimensional random walker, and the Right Panel  of Fig.~\ref{fig:2DIllu} for three sample  trajectories of this process with finite termination time).   For large and symmetric thresholds $\ell_-=\ell_+ \gg 1$, we demonstrate that   the first-passage time statistics at the negative and positive thresholds are, excluding specific parameter choices,  different, in correspondence with the general theory  in  Secs.~\ref{sec:moment} and \ref{sec:moment2}.

\subsection{Random walker on a two-dimensional lattice}\label{sec:RW2D}
Consider the spatial coordinates $X_t=(X^{(1)}_t,X^{(2)}_t)\in \left\{0,1,\ldots,n-1\right\}^2$ of a random walker that moves on a  lattice of dimensions $n\times n$ and with periodic boundary conditions.    The random walker  evolves in time according to 
\begin{equation}
X^{(i)}_t=  \left(N^{+}_{i,t}- N^{-}_{i,t}\right){\rm mod}\:  n,\quad i\in \left\{1,2\right\},
\end{equation}
where $N^{+}_{i,t}$ and  $N^{-}_{i,t}$ are counting processes with rates $k^+_i$ and $k^-_i$, respectively, and where mod refers to the modulo operation.   

We parameterise the rates  by 
\begin{equation}
k^+_1 = \frac{e^{\nu/2}}{4\cosh(\nu/2)}, \quad k^-_1 = \frac{e^{-\nu/2}}{4\cosh(\nu/2)}, \label{eq:rate1} 
\end{equation}
and 
\begin{equation}
k^+_2 = \frac{e^{\nu \rho/2}}{4\cosh(\nu \rho/2)}, \quad k^-_2 = \frac{e^{-\nu \rho/2}}{4\cosh(\nu \rho/2)}, \label{eq:rate2} 
\end{equation}
so that $k^+_1 + k^-_1  + k^+_2 + k^-_2 = 1$, and we are left with two  parameters $\nu$ and $\rho$.   With this parametrisation, the  rate of dissipation (\ref{eq:sdotDef}) reads  
\begin{equation}
\dot{s} = \nu (k^+_1-k^-_1) + \nu \rho(k^+_2-k^-_2) .   \label{eq:sdot2D}
\end{equation}

\subsection{Fluctuating current $J_t$}
We take as our time-additive observables of interest   currents of the form
\begin{equation}
J_t = (1-\Delta)(N^+_{1,t}-N^-_{1,t}) + (1+\Delta)(N^+_{2,t}-N^-_{2,t})  ,  \label{eq:Jexam}
\end{equation}
that are specified by the parameter $\Delta\in \mathbb{R}$; notice that the currents $J$  can  also  be expressed  in the canonical form (\ref{eq:generic}).    For such fluctuating currents  the statistics of $T$ are independent of the lattice length  $n$ (this is however not  the case for other time-additive observables, such as, the time that the particle spends in a certain state).

If
\begin{equation}
\Delta = \frac{\rho-1}{\rho+1}, \label{eq:DeltaS}
\end{equation}
then
\begin{equation}
J_t = \frac{2}{\nu(\rho+1)}\: S_t,
\end{equation}
so that for this choice of $\Delta$ the current is  proportional to the stochastic entropy production.

\subsection{Martingale}
The cumulant generating function $\lambda_J(a)$ of $J_t$ is  given by 
\begin{equation}
\lambda_J(a) = (e^{-a(1-\Delta)}-1)k^+_1+ (e^{a(1-\Delta)}-1)k^-_1+ (e^{-a (1+\Delta)}-1)k^+_2 +  (e^{a(1+\Delta)}-1)k^-_2, \label{eq:lambdaJ2D}
\end{equation} 
as it is the Perron root  of the tilted matrix 
\begin{eqnarray}
\fl \tilde{\mathbf{q}}_{(x_1,x_2);(y_1,y_2)}(a) &=& k^+_1  e^{-a (1-\Delta)}\delta_{(x_1+1,x_2),(y_1,y_2)} +   k^-_1  e^{a (1-\Delta)}\delta_{(x_1-1,x_2),(y_1,y_2)}  \nonumber\\ 
&& + k^+_2   e^{-a (1+\Delta)}\delta_{(x_1,x_2+1),(y_1,y_2)} + k^-_2   e^{a (1+\Delta)} \delta_{(x_1,x_2-1),(y_1,y_2)} - 1.
\end{eqnarray}
Since  the corresponding right eigenvector is given by  $\phi_a(x) =1$, the Perron martingale $M_t$ of  Eq.~(\ref{eq:Mart}) takes here the  form 
\begin{equation}
M_t = \exp\left(-aJ_t - \lambda_J(a)t\right) \label{eq:Mt2D}
\end{equation}
where $J$  and $\lambda_J(a)$ are given by the Eqs.~(\ref{eq:Jexam}) and (\ref{eq:lambdaJ2D}), respectively; notice that this is the same martingale that appears in  Appendix E of Ref.~\cite{roldan2022martingales}.     Comparing Eq.~(\ref{eq:Mart}) with (\ref{eq:Mt2D}), we observe that in this example the prefactor to the exponential is trivial.

\subsection{Splitting probabilities and moment generating functions }

\begin{figure}[h!]\centering
   { \includegraphics[width=0.39\textwidth]{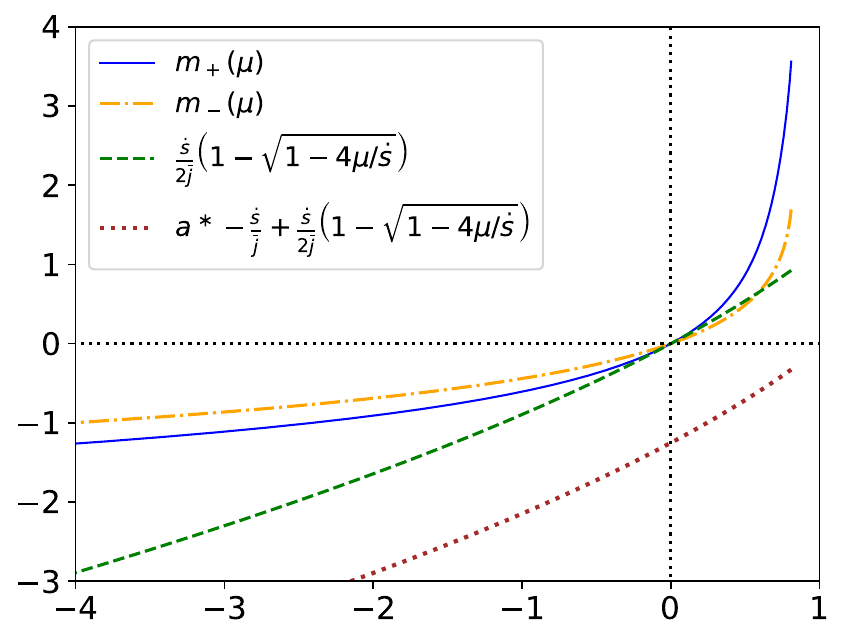}
   \put(-90,-8){\small $\mu$} 
   }
{\includegraphics[width=0.4\textwidth]{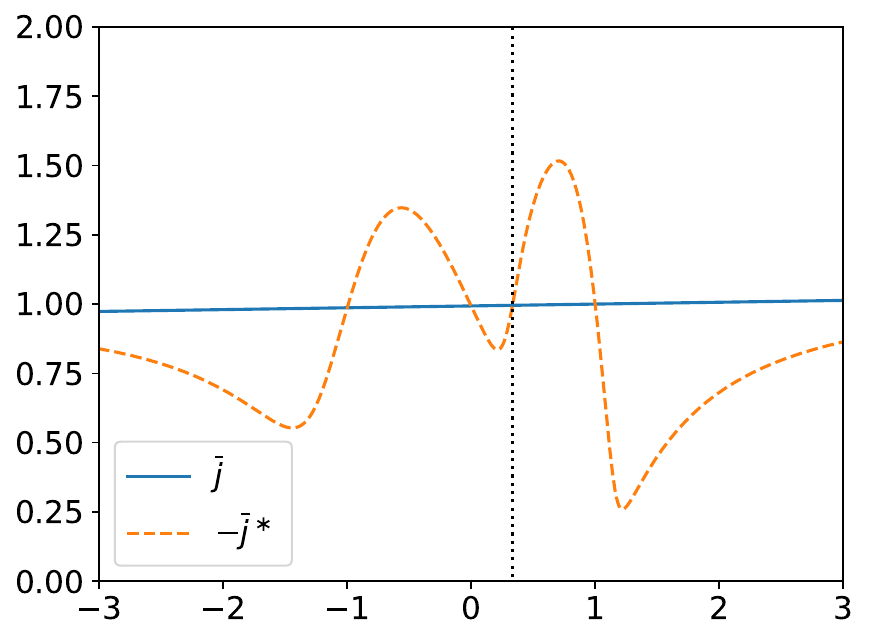}   \put(-85,-8){\small $\Delta$} }
{\includegraphics[width=0.4\textwidth]{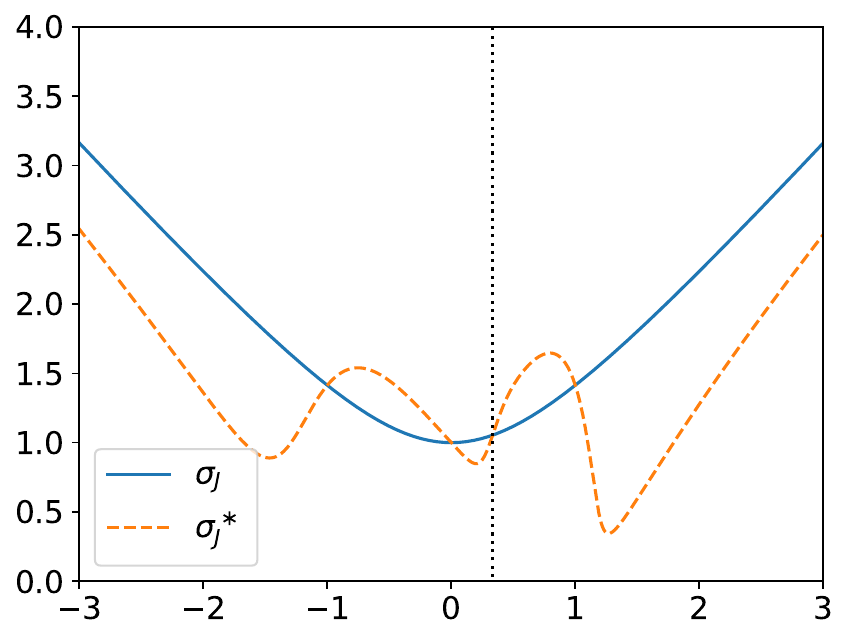} \put(-85,-8){\small $\Delta$}}
\caption{Top Left: The scaled cumulant generating functions  $m_+(\mu)$ (solid line) and $m_-(\mu)$ (dashed-dotted line)  for the first passage time $T$ of  a current $J$ in the two-dimensional random walk model of Sec.~\ref{sec:RW2D}, and comparison   with the bounds  Eqs.~(\ref{eq:bound1x}) (dashed line) and (\ref{eq:bound2x}) (dotted line), respectively.   The current $J$ is of the form Eq.~(\ref{eq:Jexam})  with $\Delta=0.7$.   The cumulant generating functions  $m_+$ and $m_-$ are obtained from Eqs.~(\ref{eq:mPx}) and (\ref{eq:mMx}), respectively, where $a^\ast$ is the nonzero solution of (\ref{eq:aAst2D}), and where $-a_-(\mu)$ and $a_+(\mu)$ are the two solutions of Eq.~(\ref{eq:tobeExpv2x}); $\dot{s}$ is  the rate of dissipation given by  (\ref{eq:sdot2D}).  
Top Right: Comparison between the average currents   $\overline{j}$ (Eq.~(\ref{eq:JMean2D}), solid line) and $\overline{j}^\ast$  (Eq.~(\ref{eq:JMean2Dx}), dashed line) in the forward process and the dual processes, respectively.         The vertical dotted line shows $\Delta=1/3$, corresponding with $J_t=2 S_t/( \nu (\rho+1))$. 
Bottom: similar plot as for the Top Right Panel, but now for the second cumulants $\sigma_J$ and $\sigma^\ast_J$ obtained from the Eqs.~(\ref{eq:sigma2J2D})  and  Eq.~(\ref{eq:sigmaJAst}), respectively.    In all panels the model parameters are $\nu=5$ and $\rho=2$.      }\label{fig:2DRandomwalkResults}   
\end{figure}    

In the present example we cannot apply the formulae  derived in  Sec.~\ref{sec:finite}, as   the  condition (\ref{eq:property}) is not satisfied.  Indeed,   $J_T\notin \left\{-\ell_-,\ell_+\right\}$ (except  when $\Delta=0$, $\Delta=1$ or $\Delta=-1$).   This is  because  the current $J_t$ has two jump sizes, $1-\Delta$ or $1+\Delta$, depending on whether the process jumps horizontally or vertically  on the 2D lattice.  
 Nevertheless, since the martingale $M_t$ has a trivial  prefactor, $\phi_a=1$, the  study of the splitting probabilities and  first-passage time statistics     simplifies. 

The statistical properties of $T$ are determined by the overshoot variables $J_T-\ell_+$ and $J_T+\ell_-$ at the positive and negative thresholds through their generating functions  
\begin{equation}
h_+(a) = \langle  e^{-a (J_T -\ell_+) }\rangle_+  \quad {\rm and} \quad  h_-(a)  = \langle  e^{-a (J_T +\ell_-) }\rangle_-. 
\end{equation}

Following an analysis similar to the one presented in Sec.~\ref{sec:finite}, we find for the splitting probability 
\begin{equation}
p_- = \frac{1 - e^{-a^\ast \ell_+} h_+(a^\ast) }{  e^{a^\ast \ell_-}h_-(a^\ast) -  e^{-a^\ast \ell_+} h_+(a^\ast) },  
\end{equation} 
where the effective affinity $a^\ast$ is  the nonzero solution of the equation $\lambda_J(a^\ast)=0$, which here reads~\cite{neri2022estimating} 
\begin{equation}
0 = (e^{-a^\ast(1-\Delta)}-1)k^+_1+ (e^{a^\ast(1-\Delta)}-1)k^-_1+ (e^{-a^\ast (1+\Delta)}-1)k^+_2 +  (e^{a^\ast(1+\Delta)}-1)k^-_2. \label{eq:aAst2D}
\end{equation} 
In the limit of large thresholds, we recover the formula (\ref{eq:pmin}) for the splitting probability $p_-$, with $a^\ast$ the nonzero solution of   Eq.~(\ref{eq:aAst2D}).

Also the generating functions $g_+$ and $g_-$ of $T$ are determined by the  statistics of the overshoot variables.     Applying Doob's optional stopping theorem to the martingale (\ref{eq:Mt2D}),  we obtain
\begin{equation}
g_+(\mu)=\frac{1}{p_+} \frac{ e^{a_+ \ell_-} h_-(a_+)  - e^{-a_-\ell_-} h_-(-a_-)}{e^{a_-\ell_++a_+\ell_-}h_+(-a_-)  h_-(a_+) - e^{-a_-\ell_--a_+\ell_+}h_-(-a_-)h_+(a_+)   } \label{eq:gPIntermedx}
\end{equation}
and 
\begin{equation}
 g_-(\mu) =\frac{1}{p_-} \frac{e^{a_-\ell_+}h_+(-a_-)- e^{-a_+ \ell_+}h_+(a_+)}{e^{a_-\ell_++a_+\ell_-}h_-(a_+)h_+(-a_-)  -  e^{-a_- \ell_--a_+\ell_+}h_-(-a_-)h_+(a_+)},
 \label{eq:gMIntermedx}
\end{equation}
where $a_+(\mu)>-a_-(\mu)$ are the roots of $\lambda_J(a)=-\mu$;  in the present model,  they are obtained from solving 
\begin{equation}
-\mu= (e^{-a(1-\Delta)}-1)k^+_1+ (e^{a (1-\Delta)}-1)k^-_1+ (e^{-a (1+\Delta)}-1)k^+_2 +  (e^{a (1+\Delta)}-1)k^-_2  \label{eq:tobeExpv2x}
\end{equation} 
with $a=a_+$ and $a=-a_-$.  
In the limit of large thresholds, we recover the Eqs.~(\ref{eq:mP}) and (\ref{eq:mM})  for the scaled cumulant generating functions $m_+$ and $m_-$.    Expanding $g_+(\mu)$ and $g_-(\mu)$ in $\mu$, we obtain   Wald's equality 
$\langle T\rangle = (p_-  \langle J_T\rangle_- + p_+  \langle J_T\rangle_+)/\overline{j}$ that also applies at finite thresholds~\cite{Wald1,Wald2}.

 The Top Left Panel of Fig.~\ref{fig:2DRandomwalkResults} plots $m_+(\mu)$ and $m_-(\mu)$  as a function of $\mu$ for the parameter choices  $\rho=2$, $\nu=5$, and $\Delta=0.7$.   The two functions $m_+(\mu)$ and $m_-(\mu)$  are different, confirming that the statistics of $T$ at both thresholds are different, even when $\ell_-=\ell_+$.    The Figure also plots the right-hand side of the inequalities (\ref{eq:bound1x}) and (\ref{eq:bound2x})  as a function of $\mu$,  expressing  thermodynamic  bounds  for $m_+(\mu)$ and $m_-(\mu)$ in terms of the rate of dissipation $\dot{s}$.   Note that the bound (\ref{eq:bound2x}) for $m_-$ does not go through the origin of the plot, and thus  (\ref{eq:bound2x}) does not imply  the  thermodynamic uncertainty relation   (\ref{eq:TM}) for first-passage times at  negative thresholds.

 \subsection{Comparing the statistics of $T$ at the positive and the negative thresholds}
From the Top Left Panel  of Fig.~\ref{fig:2DRandomwalkResults}, we observe that the slope of the two functions  $m_-(\mu)$ and $m_+(\mu)$  at $\mu=0$ are different, and thus the average first-passage times $\langle T\rangle_+$ and $\langle T\rangle_-$ are different. 
Next, we provide a more detailed comparison between the statistics of $T$ at both thresholds.    

According to the Eqs.~(\ref{eq:omean}) and (\ref{eq:ovariance}), the first  two  cumulants of $T$ at the positive thresholds, rescaled by $\ell_+$, are determined by  the average current $\overline{j}$  and the diffusivity coefficient  $\sigma^2_J$.  
Using  that $\lambda_J(a)$ is the  scaled cumulant generating function, we obtain from Eq.~(\ref{eq:lambdaJ2D}) the expressions 
\begin{equation}
\overline{j} =  (1-\Delta)   (k^+_1 - k^-_1)  +  (1+\Delta)   (k^+_2 -  k^-_2)  \label{eq:JMean2D}
\end{equation}
and 
\begin{equation}
\sigma^2_{J} = (1-\Delta)^2   (k^+_1  + k^-_1 )  + (1+\Delta)^2   (k^+_2 +  k^-_2 ), \label{eq:sigma2J2D}
 \end{equation}   
 which yield through Eqs.~(\ref{eq:omean}) and (\ref{eq:ovariance}) an explicit expression for the mean first-passage time and the variance of the first-passage time, respectively.

For the cumulants of $T$ at the negative threshold, we have the analogous formulae Eqs.~(\ref{eq:TAvoAst}) and (\ref{eq:TSigmaJEA}).    However, in this case the cumulants are determined by the  average current $\overline{j}^\ast$ and the diffusivity coefficient  $(\sigma^\ast_J)^2$ of $J$ evaluated in  the dual process that has the  rate matrix $\mathbf{q}^\ast$, given by Eq.~(\ref{eq:qDual}).  In the present model, since $\phi_a$ is a constant function, the tilted matrix of the dual process is given by  
\begin{eqnarray}
\fl \tilde{\mathbf{q}}^\ast_{(x_1,x_2);(y_1,y_2)}(a)  &=& k^+_1  e^{-(a+a^\ast) (1-\Delta)}\delta_{(x_1+1,x_2),(y_1,y_2)} +   k^-_1  e^{ (a+a^\ast)(1-\Delta)}\delta_{(x_1-1,x_2),(y_1,y_2)}  \nonumber\\ 
&& + k^+_2   e^{-(a+a^\ast) (1+\Delta)}\delta_{(x_1,x_2+1),(y_1,y_2)} + k^-_2   e^{(a+a^\ast) (1+\Delta)} \delta_{(x_1,x_2-1),(y_1,y_2)} - 1,  \nonumber\\
\end{eqnarray} 
and thus we   recover  the equality Eq.~(\ref{eq:lambdaRel})
between the scaled cumulant generating functions in the dual  and the original process.     The current rate in the dual process is thus  
\begin{eqnarray}
\fl \overline{j}^\ast 
= (1-\Delta)  \left( e^{-a^\ast(1-\Delta)}  k^+_1 - e^{a^\ast(1-\Delta)}k^-_1 \right) + (1+\Delta)  \left(e^{-a^\ast (1+\Delta)}   k^+_2 - e^{a^\ast (1+\Delta)} k^-_2 \right) \label{eq:JMean2Dx}
\end{eqnarray}
and the diffusivity coefficient  of $J$ in the dual process is given by 
\begin{eqnarray}
\fl \left(\sigma^\ast_J\right)^2
= (1-\Delta)^2 (e^{-a^\ast(1-\Delta)}   k^+_1 + e^{a^\ast(1-\Delta)} k^-_1 ) +  (1+\Delta)^2  (e^{-a^\ast (1+\Delta)}  k^+_2 + e^{a^\ast (1+\Delta)}k^-_2).    \nonumber\\ \label{eq:sigmaJAst} 
\end{eqnarray}

In the Top Right  and Bottom Panels  of Fig.~\ref{fig:2DRandomwalkResults} we compare the average current $\overline{j}$ with $-\overline{j}^\ast$ and the  diffusivity coefficient  $
\sigma_J$ with $\sigma^\ast_J$, respectively.     We observe that  there is  no clear relationship between these quantities, as the  current $-\overline{j}^\ast$  can be smaller or larger than $\overline{j}$.    This implies that  also   $\langle T\rangle_- $ can be smaller or larger than $\langle T\rangle_+$, depending on the system parameters.     In this example, for the four values  $\Delta=-1$, $\Delta=0$, $\Delta =1/3$, and $\Delta=1$   the dual process   generates up to a sign difference  the same cumulants  as the forward process.    For these values of $\Delta$ the fluctuating current $J$ satisfies the Gallavotti-Cohen-like fluctuation relation (\ref{eq:lebGol}), and therefore  $m_+=m_-$, yielding  the same $T$ statistics    at  both thresholds.   For  $\Delta=1/3$ the fluctuating current $J_t$ is proportional to $S_t$, as for this value of $\Delta$  Eq.~(\ref{eq:DeltaS})  is satisfied (notice that $\rho=2$), and hence in this case $J_t$ is an optimal current for which $\dot{s} = \overline{j}a^\ast$.  The other three values correspond with $\Delta=1$, $\Delta=0$, and $\Delta=-1$,   so that $J_t = 2 (N^+_2-N^{-}_2)$, $J_t =N^+_1-N^{-}_1 +  N^+_2-N^{-}_2$, and $J_t = 2 (N^+_1-N^{-}_1)$, respectively.  In these cases the  fluctuating current   $J_t$ denotes the position of a biased random walker on a one-dimensional lattice and $J_t$ is  proportional to the stochastic entropy  production of the corresponding  one-dimensional biased random walker.  For this reason, also at  $\Delta=1$ and $\Delta=-1$   the Gallavotti-Cohen fluctuation relation is satisfied  and consequently $m_-=m_+$, yielding the same statistics of $T$ at both thresholds.

\section{Run-and-tumble motion: first-passage properties in a diffusive nonequilibrium system and a martingale duet} \label{sec:runAndTumble}
As shown in  Sec.~\ref{sec:LDPMean}, for nonequilibrium systems with zero average currents, i.e., $\overline{j} = 0$, the splitting probabilities and  mean first-passage times exhibit in the limit of large thresholds diffusive behaviour.  A prominent example of nonequilibrium and diffusive motion is  realised by run-and-tumble motion.  This process  is characterized by directed runs that are   interrupted by tumbles during which the walker randomly reallocates its direction~\cite{cates2012diffusive}.  Run-and-tumble motion can be found in   peritrichous bacteria, such as,   Escherichia coli~\cite{berg2004coli}, that have  flagella located randomly at various positions on the cell body~\cite{lauga2016bacterial}.  In one-dimensional continuous space, the first-passage problem for a run-and-tumble particle exiting a finite interval --- analogous to the gambler's ruin problem  --- has been studied in Ref.~\cite{malakar2018steady}. Here, we consider the first-passage problem for run-and-tumble particles on a one-dimensional lattice~\cite{masoliver2017continuous, jose2022active}, which to the best of my knowledge has not been solved before.    

We  solve the first-passage problem of run-and-tumble particles exiting an interval of finite width  by  using a  {\it duet of martingales}, consisting of the family of  Perron martingales (\ref{eq:Mart})  and a second family of nonpositive martingales of the form (\ref{eq:Mart2}).  This demonstrates the use of nonpositive martingales for solving first-passage problems at finite thresholds.

\subsection{Model definition} 
\begin{figure}[h!]\centering
 \includegraphics[width=0.5\textwidth]{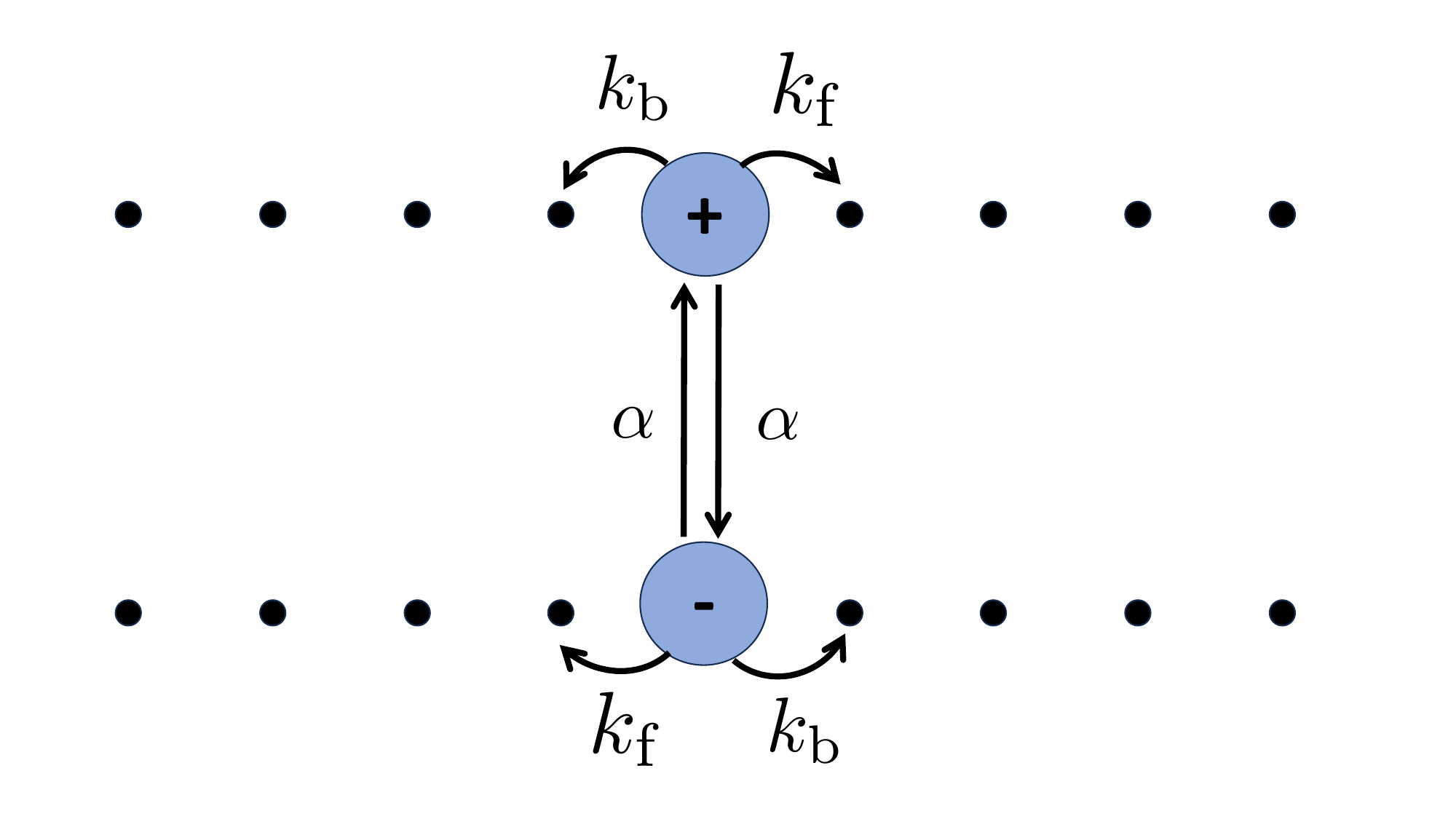}
 \newline
 { \includegraphics[width=0.37\textwidth]{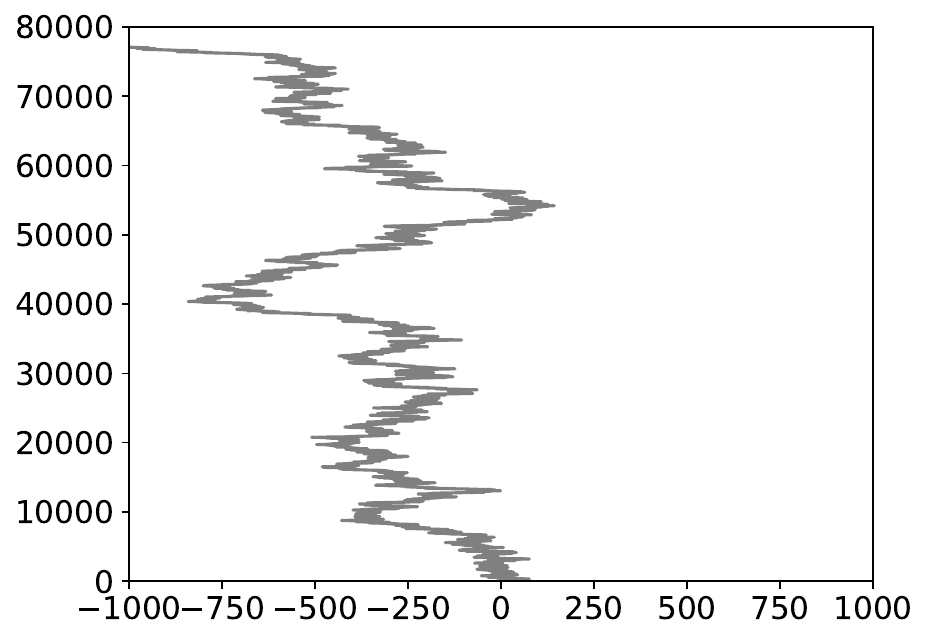}
 \put(-175,60){\small $t$} 
 \put(-80,-8){\small $Y_t$} 
 }
{\includegraphics[width=0.37\textwidth]{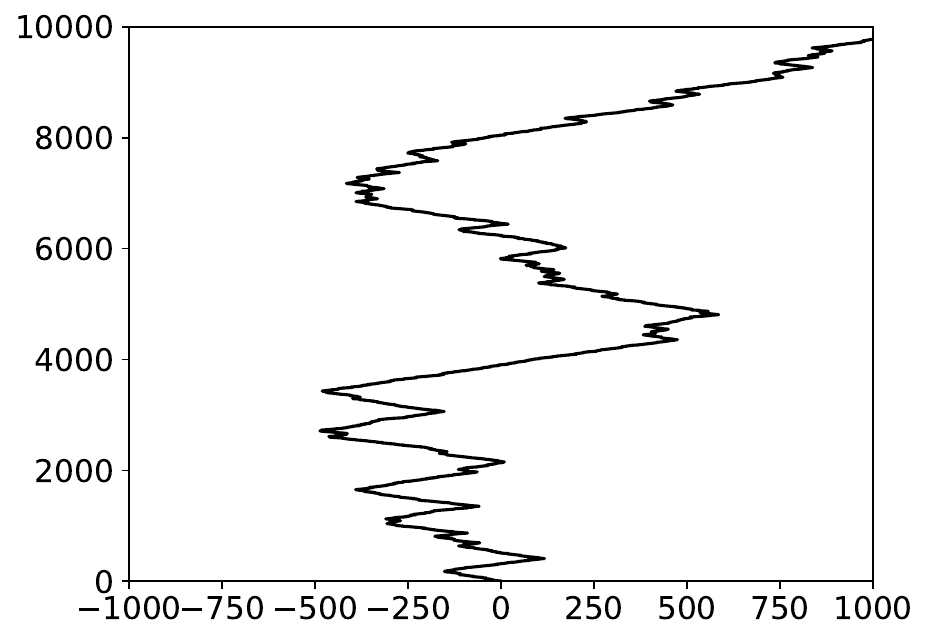}
 \put(-170,60){\small $t$} 
 \put(-80,-8){\small $Y_t$} }
\caption{ Escape problem for a one-dimensional run-and-tumble particle.     Top Panel: Illustration of the rates in the  studied model for  one-dimensional  run-and-tumble motion.   Bottom Panel:  kymographs (y-axis is time and x-axis is space)  of two trajectories corresponding with $k_{\rm f}=2$, $k_{\rm b}=1$, $\ell_-=\ell_+=1000$ and $\alpha=0.1$ (left) and $\alpha=0.01$ (right).}\label{fig:RTIllu}   
\end{figure}

Consider a particle that moves on  a one-dimensional lattice  with $n$ sites and with periodic boundary conditions forming the discrete equivalent of a ring.   The particle comes in two {\it polarisation} states that we denote  by the $+$-state and the $-$-state.   If the particle is in the $+$-state, then it hops clockwise at a rate $k_{\rm f}$ and counterclockwise at a rate $k_{\rm b}$;  if the particle is in the $-$-state, then it hops counterclockwise at rate $k_{\rm f}$ and clockwise at a rate $k_{\rm b}<k_{\rm f}$.     The particle tumbles  between the $+$ and the $-$-states at a rate $\alpha$.      We assume in what follows, without loss of generality, that $k_{\rm f}>k_{\rm b}$, so that the particle drifts on average clockwise (Fig.~\ref{fig:RTIllu} gives a graphical illustration of the model).

The Markov process $X_t=(Y_t, Z_t)$ is thus a pair with 
$Y_t\in \left\{1,2,\ldots,n\right\}$ the particle's position and $Z_t\in\left\{+,-\right\}$ the particle's polarisation.  The  nonzero entries of the rate matrix  $\mathbf{q}$ are 
\begin{equation}
\mathbf{q}_{(y,+),(y,-)} = \alpha, \quad {\rm and} \quad  \mathbf{q}_{(y,-),(y,+)} = \alpha, 
\end{equation} 
\begin{equation}
\mathbf{q}_{(y,+),(y+1,+)} = k_{\rm f},   \quad {\rm and} \quad   \mathbf{q}_{(y,+),(y-1,+)} = k_{\rm b},
\end{equation} 
\begin{equation}
\mathbf{q}_{(y,-),(y+1,-)} = k_{\rm b} ,   \quad {\rm and} \quad    \mathbf{q}_{(y,-),(y-1,-)} = k_{\rm f},
\end{equation} 
where $y\in \left\{1,2,\ldots,n\right\}$.

We assume, for simplicity, that the initial state is stationary so that 
\begin{equation}
p_{Z_0}(+) = p_{Z_0}(-) = \frac{1}{2}; \label{eq:initZ}
\end{equation}
  nonuniform  initial conditions slightly complicate the following analysis, but can be solved as well.

We take as our  current of interest the net total distance that the particle has moved along the lattice, i.e., 
\begin{equation}
J_t := N^+_t-N^-_t,\label{eq:JRT}
\end{equation}
where $N^+_t$  and  $N^-_t$ count the number of times that    $Y_{s+0^+}-Y_s = 1$  and  $Y_{s}-Y_{s+0^+} {\rm mod}\:n = 1$, respectively, for $s\in [0,t]$; since we use periodic boundary conditions, it should be understood that $Y_{s}-Y_{s+0^+} = n-1=-1$ and $Y_{s}-Y_{s+0^+}  = 1-n=1$.   In summary, the first passage problem $T$ is  the exit problem of a run-and-tumble particle out of an interval of finite width, and we illustrate some representative trajectories in the bottom panels of Fig.~\ref{fig:RTIllu}.

In this model, the  average current $\overline{j} = 0$, even though the process is not in equilibrium.   Therefore,  we are considering  a first-passage problem in a      diffusive, nonequilibrium system.      

\subsection{Martingale duet}

In this case, the $\tilde{\mathbf{q}}$ matrix has two eigenvalues (see~\ref{appendix:RT}), viz.,
\begin{eqnarray}
\fl \mu^{\pm}(a) =  \frac{ e^{-a} }{2}\left( \left(e^{2a} +1\right) (k_{\rm b}  + k_{\rm f} ) \pm  \sqrt{4\alpha^2 e^{2a} + (e^{2a}-1)^2(k_{\rm b}-k_{\rm f})^2 }  \right) 
\nonumber\\ 
-(\alpha + k_{\rm b}+k_{\rm f}) , \label{eq:muPMRT}
\end{eqnarray}
with the Perron root $\lambda_J(a) = \mu^+(a)$.   We consider the right eigenvectors associated with these  eigenvalues that depend on the polarisation state $Z_t$ only, and hence these  eigenvectors  have effectively two  components.    The right  eigenvector 
associated with the Perron root  $\mu^+(a)$ takes the form  
\begin{eqnarray}
\left(\begin{array} {c}\zeta^+_a(+)  \\ \zeta^+_a(-)  \end{array}\right) 
&=&\frac{1}{\alpha} \left(\begin{array} {c} 
\mu^{+}(a)+\alpha+ (1-e^{-a})k_{\rm b}+(1-e^{a})k_{\rm f} \\ \alpha \end{array}\right) \label{eq:eigv1}
\end{eqnarray}
and the right eigenvector associated with $\mu^-(a)$ is 
\begin{eqnarray}
\left(\begin{array} {c}\zeta^-_a(+)  \\ \zeta^-_a(-)  \end{array}\right) 
&=& \frac{1}{\alpha} \left(\begin{array} {c}
\mu^{-}(a)+\alpha+ (1-e^{-a})k_{\rm b}+(1-e^{a})k_{\rm f} \\ \alpha\end{array}\right). \label{eq:eigv2}
\end{eqnarray}

The  two eigenpairs $(\mu^+,\zeta^+)$ and  $(\mu^-,\zeta^-)$  yield a duet $(M^+,M^-)$ of martingales, viz.,    
\begin{equation}
M^+_t = \zeta^+_a(Z_t) \exp\left(-aJ_t - \mu^+(a)t\right)
\end{equation}
and 
\begin{equation}
M^-_t = \zeta^-_a(Z_t) \exp\left(-aJ_t - \mu^-(a)t\right).
\end{equation}

Notice that $M^+$ is the (positive) Perron martingale, while $M^-$ is a second  one-parameter family of martingales that can  have both positive and negative values.   
Indeed, for  $a=0$ we get  
\begin{equation}
\left(\begin{array} {c}\zeta^+_0(+)  \\ \zeta^+_0(-)  \end{array}\right) =  \left(\begin{array} {c} 1 \\ 1 \end{array}\right)  \quad {\rm and } \quad  \left(\begin{array} {c}\zeta^-_0(+)  \\ \zeta^-_0(-)  \end{array}\right) =  \left(\begin{array} {c} -1 \\ 1 \end{array}\right).
\end{equation} 
At $a=0$, the Perron root  satisfies $\mu^+(0)=0$ and  $\mu'_+(0) = 0$, consistent with $\overline{j}=0$.   On the other hand, $\mu^-(0) = -2\alpha$ and   $\mu'_-(0) = 0$.

\subsection{Splitting probability}
Using  duet of two martingales, $M^+_t$ and $M^-_t$, we derive  the expression 
\begin{eqnarray}
\fl p_-   =\frac{ 2 \alpha \left(1 + e^{-\overline{a} (\ell_{-} + \ell_{+})}\right) \ell_{+} \xi_1 \xi_2 + (1-e^{-\overline{a}  \ell_{+}}) \left(k_{\rm b} -k_{\rm f}\right) \left( [2 + \xi_1] \xi_2 - \xi_1 [2+\xi_2] e^{-\overline{a}  \ell_{-}}\right)   }{ 2 \left(1- e^{-\overline{a}  (\ell_{-} + \ell_{+})}\right) (k_{\rm f}-k_{\rm b} )(\xi_1 - \xi_2)  +
    2 \alpha \left(1 + e^{-\overline{a}  (\ell_{-} + \ell_{+})}\right) (\ell_{-} + \ell_{+}) \xi_1 \xi_2}\nonumber\\ \label{eq:pMFinalT}
\end{eqnarray}
for the splitting probability, where 
\begin{eqnarray}
\xi_1 &=& \frac{1}{\alpha} \left((1-e^{\overline{a}})k_{\rm b}+(1-e^{-\overline{a}})k_{\rm f}\right) \label{eq:xi1}
\end{eqnarray}
and 
\begin{eqnarray}
\xi_2 &=& \frac{1}{\alpha} \left((1-e^{-\overline{a}})k_{\rm b}+(1-e^{\overline{a}})k_{\rm f}\right), \label{eq:xi2}
\end{eqnarray}
and 
\begin{eqnarray}
\fl  \overline{a} = \log\left(\frac{k_{\rm b}^2 + k_{\rm f}^2 + \alpha (k_{\rm b} + k_{\rm f}) +
    \sqrt{(k_{\rm b} + k_{\rm f})(\alpha + k_{\rm b} + k_{\rm f}) [(k_{\rm b} - k_{\rm f})^2 + \alpha (k_{\rm b} + k_{\rm f})]}}{2 k_{\rm b} k_{\rm f}}\right). \nonumber\\\label{eq:a2}
\end{eqnarray}

\begin{figure}[h!]\centering
{\includegraphics[width=0.43\textwidth]{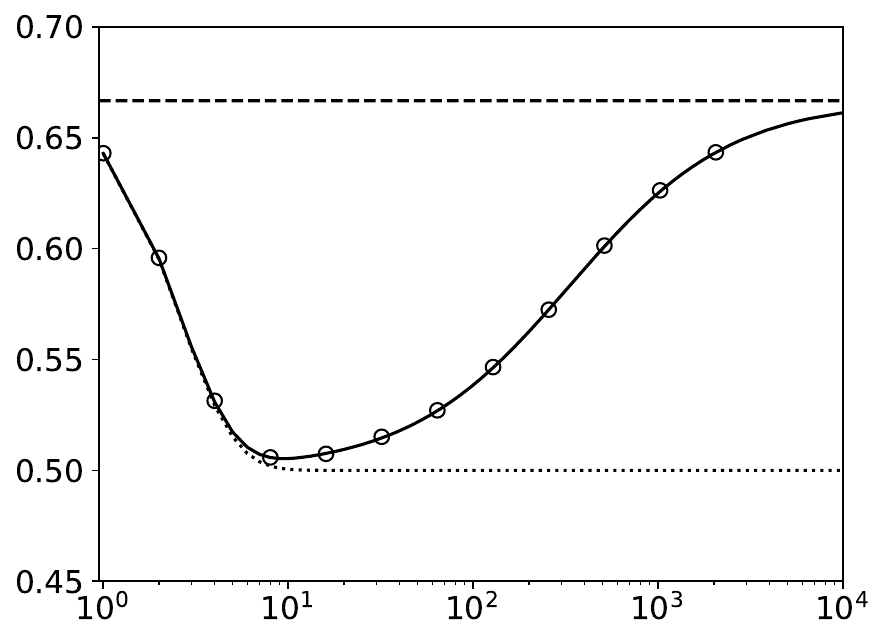}\put(-204,72){\small $p_-$} 
\put(-90,-8){\small $\ell_-$} }
\hspace{5mm}
{\includegraphics[width=0.425\textwidth]{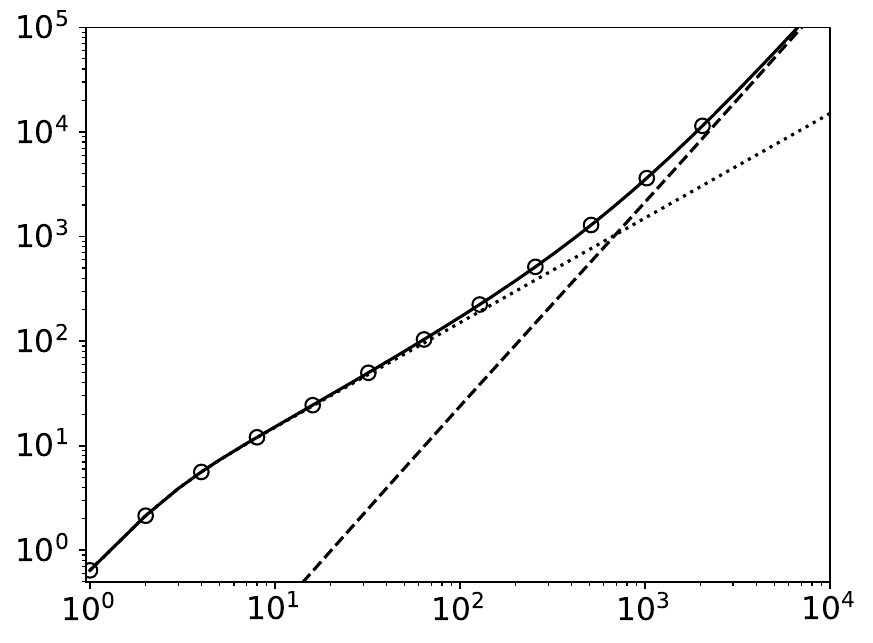}\put(-208,72){\small $\langle T\rangle$} 
\put(-90,-8){\small $\ell_-$} }
\caption{ First-passage quantities  for a run-and-tumble particle escaping an interval of width $(-\ell_-,\ell_+)$, i.e., the first-passage problem (\ref{eq:T}) for the time-additive observable  (\ref{eq:JRT}).  
Left Panel: The splitting probability $p_-$  as given by the Eq.~(\ref{eq:pMFinalT}) (solid line) is compared with empirical estimates based on     $1e+6$ simulated trajectories  (markers),  with $p_-$ at $\alpha=0$ given by the formula (\ref{eq:pMAlpha0}) (dotted line), and with   the diffusive limit for $\ell\rightarrow \infty$ given by Eq.~(\ref{eq:pMZeroO}) (which here reads $p_-=2/3$).     Right Panel:  The  mean first passage time given by formula (\ref{eq:meanTZero}) (solid line) is compared with  empirical estimates based on     $1e+6$ simulated trajectories  (markers), with the formula (\ref{eq:zeroalphaTCorrect}) for $\alpha=0$ (dotted line), and with the asymptotic formula (\ref{eq:TAsympTRT})  for $\ell\rightarrow \infty$ (dashed line).     The model parameters are:  hopping rates $k_{\rm f}=2$ and $k_{\rm b}=1$,  switching rate $\alpha = 0.001$, and the thresholds are set to $\ell_+ = 2\ell_-$.  The initial particle polarisation is given by Eq.~(\ref{eq:initZ}). }\label{fig:runTumble}   
\end{figure}  

In the left panel of Fig.~\ref{fig:runTumble}, 
we plot the splitting probability $p_-$  as a function of $\ell_-$ with the ratio $\ell_+/\ell_-=2$ fixed.  The splitting probability is a nonmonotonic function of $\ell_-$.   At small values of $\ell_-$, the result is in perfect agreement with  the splitting probability of a random walker with infinite persistence, $\alpha=0$, in which case 
\begin{equation}
p_- = \frac{1-b^{\ell_+}+b^{\ell_-}-b^{\ell_-+\ell_+}}{2 (1-  b^{
   \ell_- + \ell_+})} \label{eq:pMAlpha0}
\end{equation}
with $b=k_{\rm b}/k_{\rm f}$.   Therefore, at small values of $\ell_-$ the splitting probability  $p_-$ decreases from its initial value to $p_-\approx 1/2$.  However, at larger threshold  values  the process recovers its diffusive limit so that $p_-$ converges towards the expression given by Eq.~(\ref{eq:pMZeroO}), which in this example corresponds with $p_-=2/3$.     Hence, we conclude that run-and-tumble particles are characterised by a  nonmonotonic dependence of the splitting probability  on the threshold width.

Next, we derive the Eq.~(\ref{eq:pMFinalT}) with the duet of martingales $M^+$ and $M^-$.  Doob's optional stopping theorem applied to the Perron Martingale  yields
\begin{equation}
p_+\langle  \zeta^+_{a}(Z_T)e^{-\mu^+(a)T} \rangle_+ e^{-a \ell_+}  + p_-\langle  \zeta^+_{a}(Z_T) e^{-\mu^+(a)T}\rangle_-  e^{a \ell_-}  = \langle \zeta^+_{a}(Z_0)\rangle . \label{eq:Doob1aa}
\end{equation} 
At $a=0$ and using that $\mu^+(0)=0$, we get from Eq.~(\ref{eq:Doob1aa}) that 
\begin{equation}
p_+ + p_-  = 1  . \label{eq:splitTotal}
\end{equation}
Taking the Taylor series  in $a$  of  both sides of Eq.~(\ref{eq:Doob1aa}), and equating the coefficients in linear order in $a$, we obtain the  Eq.~(\ref{eq:pMZero}).  In the present model, Eq.~(\ref{eq:pMZero}) reads  
\begin{equation}
p_- = \frac{\ell_+ \alpha  +  (\pi_+ - 1/2) (k_{\rm f}-k_{\rm b})  }{(\ell_- + \ell_+) \alpha +  (\pi_+ - \pi_-) (k_{\rm f}-k_{\rm b})  }  ,  \label{eq:pMZeroax2xx}
\end{equation}
where  
 \begin{equation}
\pi_+ := \mathbb{P}\left(Z_T = + | X_T\geq \ell_+ \right) 
\end{equation}
and 
\begin{equation}
\pi_- := \mathbb{P}\left(Z_T = + | X_T\leq -\ell_- \right)   
\end{equation}
are the probabilities that  the particle's polarisation $Z_T$ at the termination time $T$ is in the $+$-state, conditioned on $X_T\geq \ell_+ $ and  $X_T\leq -\ell_-$, respectively.    

Note that Eq.~(\ref{eq:pMZeroax2xx}) does not provide us yet with $p_-$ as a function of the model parameters, as the probabilities  $\pi_+ $ and $\pi_- $ are not  known.    It is here that  the martingale family $M^-$ becomes useful.  Doob's optional stopping theorem applied to $M^-$   yields
\begin{equation}
p_+\langle  \zeta^-_{a}(Z_T)e^{-\mu^-(a)T} \rangle_+ e^{-a \ell_+}  + p_-\langle  \zeta^-_{a}(Z_T)e^{-\mu^-(a)T} \rangle_-  e^{a \ell_-}  = \langle \zeta^-_{a}(Z_0)\rangle. \label{eq:Doob2aa}
\end{equation}
We use Eq.~(\ref{eq:Doob2aa}) at the values of $a$ for which 
\begin{equation}
\mu^-(a)=0, 
\end{equation}
which yields two solutions, $a=\bar{a}$ and $a=-\bar{a}$, with $\bar{a}$ given by Eq.~(\ref{eq:a2}).   Substituting $a=\bar{a}$ and  $a=-\bar{a}$ in Eq.~(\ref{eq:Doob2aa}),  we get the two  equations 
 \begin{equation}
(1-p_-)\left(1+\pi_+ \xi_1 \right) e^{\bar{a}\ell_+} + p_- \left(1+\pi_- \xi_1 \right)e^{-\bar{a}\ell_-} = \frac{1}{2} \left(2+ \xi_1\right)  \label{eq:xi1Solve}
\end{equation}
and 
\begin{equation}
(1-p_-)\left(1+\pi_+ \xi_2 \right) e^{-\bar{a}\ell_+} + p_- \left(1+\pi_- \xi_2\right)e^{\bar{a}\ell_-} = \frac{1}{2} \left(2 + \xi_2\right) .\label{eq:xi2Solve}
\end{equation}  

Solving the Eqs.~(\ref{eq:pMZeroax2xx}), (\ref{eq:xi1Solve}) and (\ref{eq:xi2Solve}) for $p_-$, $\pi_-$ and $\pi_+$, we obtain an explicit expression for $p_-$, $\pi_-$ and $\pi_+$ as a function of the model rates $\alpha$, $k_{\rm b}$, $k_{\rm f}$ and the threshold values $\ell_+$ and $\ell_-$.     For the splitting probability $p_-$, we obtain Eq.~(\ref{eq:pMFinalT}), and for the probabilities $\pi_-$ and $\pi_+$ we find the formulae 
\begin{eqnarray}
\fl \pi_-  =\frac{   (k_{\rm f}-k_{\rm b}) \left(1 - e^{-\overline{a} \ell_+}\right)\left( 2 + \xi_1 -e^{-\overline{a} \ell_-} (2  + \xi_2 )\right) }{  2 \alpha \left(1 + e^{- \overline{a} (\ell_- + \ell_+)}\right) \ell_+ \xi_1 \xi_2 + (k_{\rm b}-k_{\rm f}) \left(1  -e^{-\overline{a} \ell_+}\right) \left((2 + \xi_1) \xi_2 -e^{-\overline{a} \ell_-}  \xi_1 (2 + \xi_2)\right) }  
\nonumber\\ 
\fl  - \alpha  \frac{ 2 \ell_+ (\xi_1 -\xi_2 e^{-2\overline{a}  (\ell_- + \ell_+)}) - e^{-\overline{a} \ell_- } (\ell_- + \ell_+ ) [\xi_1 (2 + \xi_2) - e^{-2\overline{a}  \ell_+}(2 + \xi_1) \xi_2  ]+ 2 e^{-\overline{a}  (\ell_- + \ell_+)} \ell_- (\xi_1 - \xi_2) }{(1-e^{-\overline{a}(\ell_-+\ell_+)})\left[2 \alpha \left(1 + e^{- \overline{a} (\ell_- + \ell_+)}\right) \ell_+ \xi_1 \xi_2 + (k_{\rm b}-k_{\rm f}) \left(1  -e^{-\overline{a} \ell_+}\right) \left((2 + \xi_1) \xi_2 -e^{-\overline{a} \ell_-}  \xi_1 (2 + \xi_2)\right)\right]}. \nonumber\\ \label{eq:piM}
\end{eqnarray}
and  
\begin{eqnarray}
\fl \pi_+  =-\frac{   (k_{\rm f}-k_{\rm b}) \left(1 - e^{-\overline{a} \ell_-}\right)\left( 2 + \xi_2 -e^{-\overline{a} \ell_+} (2  + \xi_1 )\right) }{  2 \alpha \left(1 + e^{- \overline{a} (\ell_- + \ell_+)}\right) \ell_- \xi_1 \xi_2 + (k_{\rm f}-k_{\rm b}) \left(1  -e^{-\overline{a} \ell_-}\right) \left((2 + \xi_2) \xi_1 -e^{-\overline{a} \ell_+}  \xi_2 (2 + \xi_1)\right) }  
\nonumber\\ 
\fl  - \alpha  \frac{ 2 \ell_- (\xi_2 -\xi_1 e^{-2\overline{a}  (\ell_- + \ell_+)}) - e^{-\overline{a} \ell_+ } (\ell_- + \ell_+ ) [\xi_2 (2 + \xi_1) - e^{-2\overline{a}  \ell_-}(2 + \xi_2) \xi_1  ]+ 2 e^{-\overline{a}  (\ell_- + \ell_+)} \ell_+ (\xi_2 - \xi_1) }{(1-e^{-\overline{a}(\ell_-+\ell_+)})\left[2 \alpha \left(1 + e^{- \overline{a} (\ell_- + \ell_+)}\right) \ell_- \xi_1 \xi_2 + (k_{\rm f}-k_{\rm b}) \left(1  -e^{-\overline{a} \ell_-}\right) \left((2 + \xi_2) \xi_1 -e^{-\overline{a} \ell_+}  \xi_2 (2 + \xi_1)\right)\right]}.  \nonumber\\\label{eq:piP}
\end{eqnarray}
The latter two Eqs.~(\ref{eq:piM}) and (\ref{eq:piP}) are useful for expressing $\langle T\rangle$ in terms of the model parameters, as we show in the next subsection.

\subsection{Mean first-passage time}
Taking the  Taylor series of both sides of the equality in Eq.~(\ref{eq:Doob1aa}) and equating the  coefficients  in second order in $a$ we obtain Eq.~(\ref{eq:meanTZero}), which here reads 
\begin{eqnarray}
 \langle T\rangle   &=&  \frac{(1-p_-)}{\sigma^2_J}  \left( \ell^2_+ + 2\ell_+ \pi_+ \frac{k_{\rm f}-k_{\rm b}}{\alpha}  +  \pi_+ \left(\frac{k_{\rm f}-k_{\rm b}}{\alpha}\right)^2 \right)  
\nonumber\\ 
&& 
+\frac{p_-}{\sigma^2_J }  \left( \ell^2_- - 2\ell_- \pi_- \frac{k_{\rm f}-k_{\rm b}}{\alpha} +  \pi_- \left(\frac{k_{\rm f}-k_{\rm b}}{\alpha}\right)^2 \right) 
-\frac{1}{2\sigma^2_J}\left(\frac{k_{\rm f}-k_{\rm b}}{\alpha}\right)^2  , \nonumber\\  \label{eq:meanTZerov2}
\end{eqnarray}  
where 
\begin{equation}
\sigma^2_J =  k_{\rm  b} + k_{\rm f}+ \frac{(k_{\rm  b} - k_{\rm  f})^2}{\alpha} 
\end{equation} 
is the diffusivity coefficient.   Substituting the expressions for $p_-$, $\pi_-$, and $\pi_+$ as a function of the model parameters [i.e., the Eqs.~(\ref{eq:pMFinalT}), (\ref{eq:piM}), and (\ref{eq:piP}) derived in the previous section] into (\ref{eq:meanTZerov2}), we obtain an  expression for $\langle T\rangle$ as a function of $k_{\rm f}$, $k_{\rm b}$, $\alpha$, $\ell_+$, and~$\ell_-$.   

The right panel of Fig.~(\ref{fig:runTumble}) plots $\langle T\rangle$ as a function of $\ell_-$ for a fixed ratio $\ell_+/\ell_-=2$.    Just as for the splitting probability, we can observe two regimes,  corresponding with  directed and diffusive motion.    At small values of $\ell_-$,  the mean first-passage time is well described by the expression for $\langle T\rangle$ at $\alpha=0$, viz., 
\begin{eqnarray}
\langle T\rangle = \frac{\ell_+ + \ell_- }{2(k_{\rm f}-k_{\rm b})} \left(\frac{ 1- b^{\ell_+} -  b^{\ell_-}+b^{\ell_++\ell_-}}{1-b^{\ell_++\ell_-}}  \right),
\label{eq:zeroalphaTCorrect}
\end{eqnarray}
with $b=k_{\rm b}/k_{\rm f}$,
which predicts a linear growth of $\langle T\rangle$ with $\ell_-$.   On the other hand, in the limit of large $\ell_-$, we recover the asymptotic behaviour predicted by Eq.~(\ref{eq:meanTAsymptoZero}), which here for $\ell_+ = 2\ell_-$ grows quadratically in $\ell_-$, viz., 
\begin{equation}
    \langle T\rangle = \frac{2\ell^2_-}{\sigma^2_J}  + O(\ell_-). \label{eq:TAsympTRT}
\end{equation}

\section{Discussion}  \label{sec:discussion}
First,  in Sec.~\ref{sec:summary} we summarise this Paper's results, and then in Sec.~\ref{sec:gen}  we discuss a few open problems and possible generalisations.  
\subsection{Summary}\label{sec:summary}
We have developed  a method  based on martingales for studying  the first-passage properties of   time-additive observables in  Markov jump processes.    The martingale approach is versatile, as there exist several one-parameter families of  martingales, one for each of the eigenpairs of the tilted matrix $\tilde{\mathbf{q}}$ (see Eq.~(\ref{eq:Mart2})).     When used in conjunction with Doob's optional stopping theorem, these martingales provide us with sets of linear equations that can be solved for the splitting probability $p_-$, and the generating functions $g_+(\mu)$ and $g_-(\mu)$.    

\subsubsection{Universal first-passage statistics for large thresholds.}
In the limit of large thresholds,  the first-passage statistics of a time-additive observable $O$ are determined by  the Perron martingale   (\ref{eq:Mart}) associated with the Perron root of  the tilted matrix $\tilde{\mathbf{q}}$.    Using the Perron martingale, we  have found  three qualitatively distinct cases for the first-passage statistics, depending on the properties of the scaled cumulant generating function $\lambda_O$ that describes the large deviations of $O$.   These three cases  correspond with  different  scaling properties of the splitting probability $p_-$  with $\ell_-$:
\begin{itemize}
    \item {\it Exponentially suppressed events}: if $\lambda_O(a)$ has a nonzero root, $\lambda_O(a^\ast) = 0$, and $\overline{o}>0$, then  the events at the negative threshold are exponentially suppressed, i.e.,
\begin{equation}
\lim_{\ell_{\rm min}\rightarrow \infty}  \frac{|\ln  p_-|}{\ell_-} = a^\ast  .  \label{eq:pminR}
\end{equation}  
  The  dynamics of $O_t$ is biased towards the positive threshold.   Correspondingly, 
the first two cumulants of $T$ at the positive threshold are given by 
    \begin{equation}
\lim_{\ell_{\rm min}\rightarrow \infty}\frac{\langle T\rangle_+}{ \ell_+} = \frac{1}{\overline{o}  },\label{eq:TAvoAstR}
\end{equation}
and 
\begin{equation}
\lim_{\ell_{\rm min}\rightarrow \infty}\frac{\langle T^2\rangle_+-\langle T\rangle^2_+}{\ell_+}  = \frac{\sigma_O^2}{\overline{o}^3} . \label{eq:TSigmaJEAR}
\end{equation}
Interestingly, the cumulants  at the negative threshold are distinct from those at the positive threshold and given by 
\begin{equation}
\lim_{\ell_{\rm min}\rightarrow \infty}\frac{\langle T\rangle_-}{ \ell_-} = \frac{1}{|\overline{o}^\ast|  },\label{eq:TAvoAstRN}
\end{equation}
and 
\begin{equation}
\lim_{\ell_{\rm min}\rightarrow \infty}\frac{\langle T^2\rangle_--\langle T\rangle^2_-}{\ell_-}  = \frac{(\sigma^\ast_O)^2}{|\overline{o}^\ast|^3} , \label{eq:TSigmaJEARN}
\end{equation}
where $\overline{o}^\ast$ and $(\sigma^\ast_O)^2$ are the average rate and diffusivity coefficient of $O$ in the dual process with a rate matrix $\mathbf{q}^\ast$ as defined in (\ref{eq:qDual}).   Higher order cumulants are determined by the scaled cumulant generating functions given by Eqs.~(\ref{eq:mP}) and (\ref{eq:mM}).

 Examples of time-additive observables that fall in the present category are fluctuating currents with a nonzero average rate, such as, the currents    in Sec.~\ref{sec:randomwalker} for a random walker on a lattice.

The fact that the statistics at both thresholds are distinct is surprising.    Indeed,  asymptotically the process $O_t$ is well approximated by a biased diffusion process, and for a one-dimensional biased diffusion  \cite{sarmiento2024area} or a  biased random walker~\cite{neri2022universal, roldan2022martingales} the statistics at both thresholds are identical.   Nevertheless, we find that the statistics at both thresholds are different.  Hence,  approximating $O_t$ by a  biased-diffusion process   does not work for the statistics of $T$ at large negative thresholds.

    \item {\it Super-exponentially suppressed events}: if $\lambda_O(a)$ does not have a nonzero root, then the events at the negative thresholds are super-exponentially suppressed so that
\begin{equation}
\lim_{\ell_{\rm min}\rightarrow \infty}  \frac{|\ln  p_-|}{\ell_-} = \infty  .  \label{eq:pminRR}
\end{equation} 
In this case, the process is strongly biased towards the positive threshold, and the first two 
cumulants of $T$ are given by Eqs.~(\ref{eq:TAvoAstR}) and (\ref{eq:TSigmaJEAR}).   Examples are observables that are nonnegative, such as, the time spent in a certain state.   The scaled cumulant generating function of $T$  is given by Eq.~(\ref{eq:mP2}).

    \item {\it Sub-exponentially suppressed events:} in this case the process $O_t$ has no bias and the large threshold dynamics is diffusive.  Examples are fluctuating currents with zero average rate, such as the position of a run-and-tumble particle (see Sec.~\ref{sec:runAndTumble}).
The probability that the process terminates at the negative threshold is given by 
\begin{equation}
p_- =  \frac{\ell_+}{\ell_-+\ell_+}   + \mathcal{O}\left(1/\ell_{\rm min}\right) 
\end{equation} 
and the mean first-passage time is 
\begin{equation}
\langle T\rangle =   \frac{\ell_+\ell_-}{\sigma^2_O} + \mathcal{O}\left(\ell_+,\ell_-\right) . \label{eq:meanTAsymptoZeroR}
\end{equation}

\end{itemize}

   \subsubsection{Role of the effective affinity.}
If  $O_t$ is a fluctuating current, then $a^\ast$ is called the  {\it effective affinity}.    The effective affinity extends properties of thermodynamic affinities in systems with uncoupled currents to the case of coupled currents~\cite{raghu2024effective}, which makes it a physically relevant quantity.       Notably, it was shown in \cite{raghu2024effective}  that (i) the effective affinity determines the direction of the fluctuating current, as $a^\ast\overline{j}>0$; (ii) the effective affinity captures a portion of the total rate of dissipation in  the process $X_t$ as $a^\ast \overline{j}\leq \dot{s}$;  (iii) if the process $X_t$ is described by a set of uncoupled currents, then the effective affinity equals the thermodynamic affinity; (iv) the effective affinity is the (asymptotic) exponential decay constant of the distribution of current infima, as also follows from (\ref{eq:pmin}).   

Furthermore, in this Paper we have shown that the effective affinity $a^\ast$ determines the scaled cumulants of the first-passage times $T$ at the unlikely threshold through the dual  process $\mathbf{q}^\ast$, see Eqs.~(\ref{eq:TAvoAstRN}) and (\ref{eq:TSigmaJEARN}).    This further highlights the relevance of the effective affinity for nonequilibrium physics.

\subsubsection{Solving first-passage problems at finite thresholds.}
Solving first-passage problems at finite thresholds is more complicated due to the dependence on the initial state $X_0$,  the final state $X_T$, and the overshoot $O_T-\ell_+$ (or $O_T+\ell_-$).     

In Sec.~\ref{sec:finite} we have  considered first-passage problems for which  the random variables $(X_T,O_T)$ are  deterministic  if they are   conditioned on termination at either the positive ($O_T\geq \ell_+$) or negative threshold ($O_T\leq -\ell_-$).    In this case,  the Perron martingale suffices to solve the problem, and we have  derived explicit expressions for $p_-$, $g_-$, and $g_+$ in terms of the Perron root $\lambda_O(a)$ of the tilted matrix $\tilde{\mathbf{q}}$ and  its corresponding right eigenvector $\phi_a$.  These results demonstrate the effect of initial conditions on the first-passage statistics at finite thresholds.

However, in general $(X_T,O_T)$ are nondeterministic when conditioned on termination at one of the two thresholds, and in such situations we need to determine the statistics of $(X_T,O_T)$ for the derivation of  $p_-$, $g_-$, and $g_+$.    Therefore it is not sufficient to consider the Perron martingale.  Nevertheless, using all the  one-parameter families of martingales given by Eq.~(\ref{eq:Mart2}), we can  solve such more complicated first-passage problems with martingales.    We have demonstrated this with  the example of a run-and-tumble particle in 
Sec.~\ref{sec:runAndTumble}.   In this case we have used a duet of two one-parameter families of martingales, one being the Perron martingale and the second one being a nonpositive martingale that is associated with a second eigenvalue of the tilted matrix.   Using this martingale duet, we have found a set of linear equations that we have  solved for $p_-$  and $\langle T\rangle$, finding explicit expressions for these latter two quantities.

\subsection{Extensions and open problems}\label{sec:gen}
In this Paper, we have focused on  time-additive observables in time-homogeneous Markov jump processes defined on a finite set $\mathcal{X}$.    Generalising the theory to other setups is certainly possible, although it requires a couple of technical modifications that we briefly discuss here.  

The main difficulty in extending the theory to Markov processes on sets $\mathcal{X}$ of infinite cardinality is to analyse  the  corresponding spectral problem  of the tilted generator.   For Markov processes defined on finite sets, the tilted generator is a matrix, and therefore the spectral problem of the tilted generator is a matrix diagonalisation  problem, see Eq.~(\ref{eq:spectralProblem}).  Instead, for Markov processes defined on sets of infinite cardinality, such as driven diffusions~\cite{tsobgni2016large,fischer2018large, proesmans_large-deviation_2019,du2023large}, the tilted generator is a linear operator that acts on a function space, and solving the  spectral problem of this operator is significantly more difficult than diagonalising a matrix.

We can also relax the stationarity assumption  by considering  periodically driven Markov jump processes.   In this case, a spectral problem akin to (\ref{eq:spectralProblem}) may be constructed based on Floquet theory~\cite{barato2018current}.

In Sec.~\ref{sec:runAndTumble} we have shown that a  duet of two martingales of the form (\ref{eq:Mart2})  solves the first-passage problem of a run-and-tumble particle out of an interval of finite width.   Clearly, this approach is extendable to more complicated first-passage problems by considering three or more martingale families of the form (\ref{eq:Mart2}).   This raises the question of which first-passage problems are exactly solvable with martingales, and whether this class contains  first-passage problems that are not solvable with  difference equations or partial differential equations~\cite{redner2001guide}.

In Sec.~\ref{sec:symm} we have derived the first-passage time fluctuation symmetry (\ref{eq:symm}) for fluctuating currents that satisfy the  Gallavotti-Cohen-like fluctuation relation (\ref{eq:lebGol}).   An example of a current satisfying this latter is the fluctuating entropy production $S_t$.   
Using time-reversal arguments,  Refs.~\cite{neri2017statistics, roldan2022martingales} derived the stronger result 
\begin{equation}
\langle T^n_S\rangle_-  =  \langle T^n_S\rangle_+,  
\end{equation}  
for $n\in \mathbb{N}$ and for finite thresholds.  
Although the derivations in \cite{roldan2022martingales} are convincing, it is not  clear how this result can be derived from Doob's optional stopping theorem applied to the Perron martingale,   as we require knowledge about the statistics of $\phi_a(X_T)$.     This reveals that some interesting properties are still to be learned about the stochastic entropy production. 

\section*{Acknowledgements}
The author is grateful to Adarsh Raghu  for   insightful discussions, and Lennart Dabelow and Rosemary Harris for useful comments.

\appendix

\section{Martingality of $\exp(-S_t)$}\label{app:entropy}
We show that for $J_t=S_t$, the fluctuating entropy production as defined in (\ref{eq:St}), it holds that 
\begin{equation}
\lambda_S(1) = 0 \quad {\rm and} \quad \phi_1(x) = 1.  \label{eq:appRes}
\end{equation}
Therefore,   the Perron martingale (\ref{eq:Mart}) of $S_t$ takes for $a=a^\ast$ the form
\begin{equation}
M_t = \exp(-S_t).  
\end{equation}
 
The  equalities in (\ref{eq:appRes}) follow from the fluctuation symmetry  
\begin{equation}
\tilde{\mathbf{q}}_{xy}(1-a) = \frac{1}{p_{\rm ss}(x)} \tilde{\mathbf{q}}^{\rm T}_{xy}(a) p_{\rm ss}(y), \label{eq:fluctSymm}
\end{equation}  
where $\tilde{\mathbf{q}}^{\rm T}$ is the transpose of $\tilde{\mathbf{q}}$. 
Indeed, from (\ref{eq:St}) it follows that 
\begin{equation}
c_{xy} = \ln \frac{\mathbf{q}_{xy}\: p_{\rm ss}(x)}{\mathbf{q}_{yx}\: p_{\rm ss}(y)} .
\end{equation}
Using these values of $c_{xy}$ in the definition (\ref{eq:qtilde}) of $\tilde{\mathbf{q}}$, we obtain 
\begin{equation}
\tilde{\mathbf{q}}_{xy}(1-a) =  \mathbf{q}_{xy} \exp\left((a-1) \ln \frac{\mathbf{q}_{xy}p_{\rm ss}(x)}{\mathbf{q}_{yx}p_{\rm ss}(y)}\right)  = \frac{1}{p_{\rm ss}(x)} \tilde{\mathbf{q}}^{\rm T}_{xy}(a) p_{\rm ss}(y). 
\end{equation} 
Setting $a=0$ in the above equation and using that $\tilde{\mathbf{q}}(0) = \mathbf{q}$, we find 
\begin{equation}
\tilde{\mathbf{q}}_{xy}(1) = \frac{1}{p_{\rm ss}(x)} \mathbf{q}^{\rm T}_{xy} p_{\rm ss}(y). \label{eq:eval}
\end{equation} 
  As $p_{\rm ss}$ is the stationary distribution of $\mathbf{q}_{xy}$, it holds that 
\begin{equation}
\sum_{y}\mathbf{q}^{\rm T}_{xy} p_{\rm ss}(y)= \sum_{y}  p_{\rm ss}(y)\mathbf{q}_{yx} = 0
\end{equation}
and thus by Eq.~(\ref{eq:eval}) also that 
\begin{equation}
\sum_{y}\tilde{\mathbf{q}}_{xy}(1) p_{\rm ss}(y) = 0. 
\end{equation}
This shows that $\lambda_S(1) = 0$ and $\phi_1 = 1$, concluding the derivation of (\ref{eq:appRes}).    Note that for $a=1$ the vector $\phi_a(x)$ is independent of $x$ due to the fluctuation symmetry (\ref{eq:fluctSymm}), but in general $\phi_a(x)$ does depend on $x$.

\section{Scaled cumulant generating function for the Right Panel of Fig.~\ref{fig:cases}}\label{App:rankone}
We derive the scaled cumulant generating function for the time that a random walker spends in a state on a periodic lattice of three sites.  This Markov jump process $X_t\in \left\{1,2,3\right\}$ has the $\mathbf{q}$-matrix
\begin{equation}
\mathbf{q} = \left(\begin{array}{ccc} -2& 1& 1 \\ 1 &-2&1 \\ 1 & 1 & -2 \end{array}\right), 
\end{equation}
and we consider the time-additive observable 
\begin{equation}
O_t = N^1_t,
\end{equation}
measuring the time spent in $X_t=1$.   The tilted matrix is thus
\begin{equation}
\tilde{\mathbf{q}} = \left(\begin{array}{ccc} -2-a& 1& 1 \\ 1 &-2&1 \\ 1 & 1 & -2 \end{array}\right),
\end{equation}
and the Perron root equals 
\begin{equation}
\lambda_O(a) = \frac{1}{2} \left(-3 - a + \sqrt{9 + 2 a + a^2}\right). \label{eq:lambdaOTime}
\end{equation}
This latter is the function plotted in the Right Panel of Fig.~\ref{fig:cases}.

\section{Eigenvalues and right eigenvectors of the tilted matrix of a run-and-tumble random walk process}\label{appendix:RT}

The nonzero entries of the tilted  $\mathbf{q}$-matrix are given by   
\begin{equation}
\tilde{\mathbf{q}}_{(y,+),(y,-)} = \alpha, \quad {\rm and} \quad \tilde{\mathbf{q}}_{(y,-),(y,+)} = \alpha, 
\end{equation} 
\begin{equation}
\tilde{\mathbf{q}}_{(y,+),(y+1,+)} = k_{\rm f} e^{-a}, \quad {\rm and} \quad  \tilde{\mathbf{q}}_{(y,+),(y-1,+)} = k_{\rm b} e^{a},
\end{equation}  
\begin{equation}
\tilde{\mathbf{q}}_{(y,-),(y+1,-)} = k_{\rm b} e^{-a}, \quad {\rm and} \quad  \tilde{\mathbf{q}}_{(y,-),(y-1,-)} = k_{\rm f} e^{a}. 
\end{equation}  

Considering right eigenvectors of the form  
\begin{equation}
\zeta_a(y,z) = \left\{\begin{array}{ccc} \zeta_{a}(+), &{\rm if }&z=+,  \\\zeta_{a}(-), &{\rm if }&z=-, \end{array}\right.
\end{equation}
we find that the $\zeta_{a}(\pm)$ satisfy
\begin{equation}
\mu\zeta_a(+) =   (k_{\rm f} e^{-a}+k_{\rm b} e^{a} - (\alpha + k_{\rm b}+k_{\rm f}) )\zeta_a(+)  + \alpha \zeta_a(-)   \label{eq:1aa}
\end{equation}
and 
\begin{equation}
\mu \zeta_a(-) =   (k_{\rm f} e^{a}+k_{\rm b} e^{-a}- (\alpha + k_{\rm b}+k_{\rm f}))\zeta_a(-)  + \alpha \zeta_a(+), \label{eq:2aa}
\end{equation} 
where $\mu$  is an eigenvalue and $\zeta_a(+)$ and $\zeta_a(-)$ are the  two entries of  the eigenvector associated with $\mu$.    Solving the Eqs.~(\ref{eq:1aa}) and (\ref{eq:2aa}) we find the two solutions $\mu=\mu^-$ and $\mu=\mu^+$  with 
\begin{eqnarray}
 \mu^{\pm}(a) =  \frac{ e^{-a} }{2}\left( \left(e^{2a} +1\right) (k_{\rm b}  + k_{\rm f} ) \pm  \sqrt{4\alpha^2 e^{2a} + (e^{2a}-1)^2(k_{\rm b}-k_{\rm f})^2 }  \right) 
\nonumber\\ 
-(\alpha + k_{\rm b}+k_{\rm f}) . \label{eq:eigenvaluesA}
\end{eqnarray}
The right eigenvector associated with $\mu^+$ is 
\begin{eqnarray}
\left(\begin{array} {c}\zeta^+_a(+)  \\ \zeta^+_a(-)  \end{array}\right) &=&  \left(\begin{array} {c} \frac{e^{-a}}{2\alpha}  \left((e^{2a}-1)(k_{\rm b}-k_{\rm f}) +\sqrt{4\alpha^2 e^{2a} + (e^{2a}-1)^2(k_{\rm b}-k_{\rm f})^2} \right)   \\ 1 \end{array}\right), 
\nonumber \\ 
\end{eqnarray}
and the right eigenvector associated with $\mu^-$ is 
\begin{eqnarray}
\left(\begin{array} {c}\zeta^-_a(+)  \\ \zeta^-_a(-)  \end{array}\right) &=&  \left(\begin{array} {c} \frac{e^{-a}}{2\alpha}  \left((e^{2a}-1)(k_{\rm b}-k_{\rm f}) -\sqrt{4\alpha^2 e^{2a} + (e^{2a}-1)^2(k_{\rm b}-k_{\rm f})^2} \right)  \\ 1 \end{array}\right) . \nonumber\\
\end{eqnarray}

\section*{References}
\bibliographystyle{ieeetr} 
\bibliography{bibliography}

\end{document}